\documentclass[12pt]{article}

\newenvironment{wileykeywords}{\textsf{Keywords:}\hspace{\stretch{1}}}{\hspace{\stretch{1}}\rule{1ex}{1ex}}

\usepackage{amsmath,amssymb}
\usepackage{graphicx}
\usepackage{color}
\usepackage{dcolumn}
\usepackage{bm}
\usepackage[numbers,super,comma,sort&compress]{natbib}
\usepackage[version=3]{mhchem}
\usepackage{verbatim}
\usepackage{dcolumn}
\usepackage{hyperref}
\usepackage[normalem]{ulem}
\usepackage{algorithm}
\usepackage{algpseudocode}
\algdef{SE}[DOWHILE]{Do}{doWhile}{\algorithmicdo}[1]{\algorithmicwhile\ #1}

\newcommand{\p}[1]{\left(#1\right)}
 
\usepackage[disable]{todonotes} 				
\newcommand*{\blauw}[1]{#1}
\newcommand*{\groen}[1]{#1}
\newcommand*{\blauwr}[1]{#1}

\newcommand*{\noter}[1]{}					

\definecolor{background-color}{gray}{0.98}

\title{Molecular enhanced sampling with autoencoders: On-the-fly collective variable discovery and accelerated free energy landscape exploration}

\author{Wei Chen\thanks{Department of Physics, University of Illinois at Urbana-Champaign, 1110 West Green Street, Urbana, IL 61801, USA.} \, and Andrew L. Ferguson\thanks{Correpsonding author. Email: alf@illinois.edu. Phone: (217) 300-2354. Fax: (217) 333-2736.} \thanks{Department of Materials Science and Engineering, University of Illinois at Urbana-Champaign, 1304 W Green Street, Urbana, IL 61801, USA.} \thanks{Department of Chemical and Biomolecular Engineering, University of Illinois at Urbana-Champaign, 600 South Mathews Avenue, Urbana, IL 61801, USA.}  \thanks{Department of Physics, University of Illinois at Urbana-Champaign, 1110 West Green Street, Urbana, IL 61801, USA.}}

\begin{document}

\maketitle

\begin{abstract}
\noindent Macromolecular and biomolecular folding landscapes typically contain high free energy barriers that impede efficient sampling of configurational space by standard molecular dynamics simulation. Biased sampling can artificially drive the simulation along pre-specified collective variables (\groen{CVs}), but success depends critically on the availability of good CVs associated with the important collective dynamical motions. Nonlinear machine learning techniques can identify such CVs but typically do not furnish an explicit relationship with the atomic coordinates necessary to perform biased sampling. In this work, we employ auto-associative artificial neural networks (``autoencoders'') to learn nonlinear CVs that are explicit and differentiable functions of the atomic coordinates. Our approach offers substantial speedups in exploration of configurational space, and is distinguished from exiting approaches by its capacity to simultaneously discover and directly accelerate along data-driven CVs. We demonstrate the approach in simulations of alanine dipeptide and Trp-cage, and have developed an open-source and freely-available implementation within OpenMM.
\end{abstract}

\begin{wileykeywords}
accelerated sampling, artificial neural networks, nonlinear dimensionality reduction, protein folding, molecular dynamics simulation
\end{wileykeywords}

\clearpage


\begin{figure}[h]
\centering
\colorbox{background-color}{
\fbox{
\begin{minipage}{1.0\textwidth}
\begin{center}
\includegraphics[width=50mm,height=50mm]{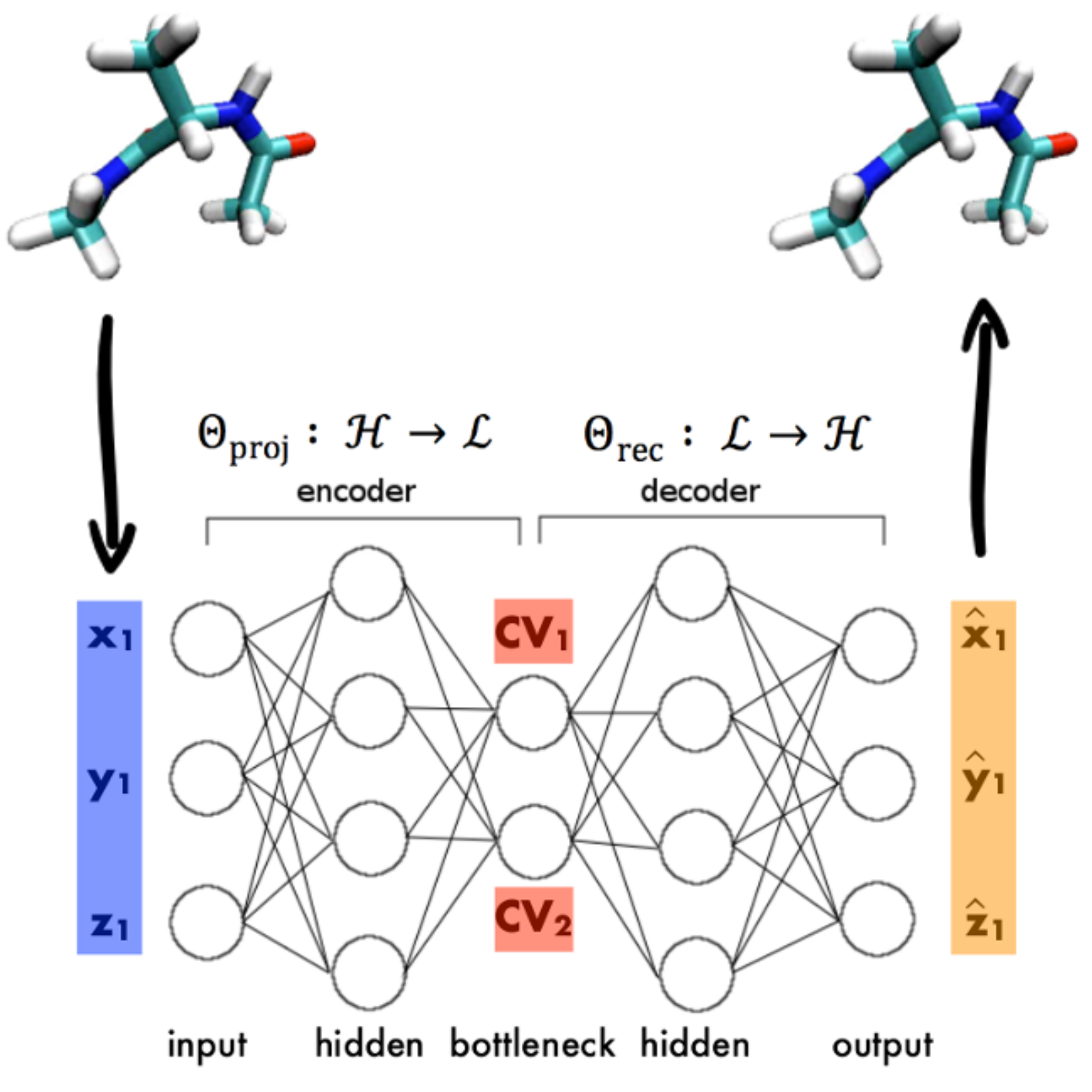}
\end{center}
Biased sampling of macromolecular and biomolecular folding requires the availability of good collective variables along which to drive sampling. We employ a variant of artificial neural networks known as autoencoders to discover data-driven collective variables from molecular simulation trajectories. Importantly, these variables are explicit and differentiable functions of the atomic coordinates that enables simultaneous collective variable discovery and accelerated sampling directly in the data-driven coordinates.
\end{minipage}
}}
\end{figure}

  \makeatletter
  \renewcommand\@biblabel[1]{#1.}
  \makeatother

\bibliographystyle{jcc}

\renewcommand{\baselinestretch}{1.5}
\normalsize

\clearpage


\section{\sffamily \Large Introduction} \label{sec:intro}

The predictive capacity of molecular dynamics (\groen{MD}) calculations on large biomolecules is limited by a disparity between the large time scales for conformational motions and the short periods accessible to simulation \cite{hashemian2013modeling,bernardi2015enhanced,karplus1990molecular}. This metastable trapping of simulations behind large free energy barriers is a consequence of rough free energy surfaces (\groen{FES}) and leads to incomplete sampling of the thermally accessible phase space \cite{hashemian2013modeling,bernardi2015enhanced,rohrdanz2013discovering,abrams2013enhanced}. Comprehensive sampling of all biologically relevant system states is required to compute converged thermodynamic averages, sample the structurally and functionally relevant molecular configurations, and explore large-scale collective motions critical to biological function such as allostery, substrate gating, and ligand binding \cite{hashemian2013modeling,bernardi2015enhanced,karplus1990molecular}. 


To confront this limitation, a plethora of enhanced sampling techniques have been developed to accelerate barrier crossing in MD simulation\cite{rohrdanz2013discovering,abrams2013enhanced}. These methodologies can be broadly partitioned into two categories: tempering / generalized ensemble techniques, and collective variable (\groen{CV}) biasing \cite{abrams2013enhanced}. We distinguish enhanced sampling methods to perform accelerated exploration of configurational space from path-based techniques such as transition path sampling (\groen{TPS}) \cite{dellago1998transition,bolhuis2002transition,rogal2008multiple,moroni2004rate,van2003novel}, string methods \cite{weinan2002string,weinan2005finite,jonsson1998nudged,sheppard2008optimization}, and forward flux sampling (\groen{FFS}) \cite{borrero2007reaction,allen2009forward} that quantify transition pathways and rates between predefined system states. Tempering techniques include simulated annealing \cite{brooks1995optimization}, multicanonical algorithm \cite{berg1992multicanonical}, replica exchange \cite{hansmann1997parallel,sugita1999replica}, and Hamiltonian exchange\cite{sugita2000replica,mitsutake2001generalized,fukunishi2002hamiltonian,okamoto2004generalized,liu2006hydrophobic}, which modify the system temperature and/or Hamiltonian to accelerate barrier crossing\cite{abrams2013enhanced}. Collective variable biasing techniques accelerate system dynamics along pre-selected CVs\cite{abrams2013enhanced,rohrdanz2013discovering}, and include non-Boltzmann sampling approaches such as umbrella sampling (\groen{US}) \cite{torrie1977nonphysical}, hyperdynamics \cite{voter1997hyperdynamics}, conformational flooding \cite{grubmuller1995predicting}, metadynamics \cite{laio2002escaping,huber1994local,barducci2008well}, adiabatic free energy dynamics (\groen{AFED})\cite{rosso2002use}, temperature accelerated molecular dynamics (\groen{TAMD}) \cite{maragliano2006temperature} / driven adiabatic free energy dynamics (\groen{d-AFED}) \cite{abrams2008efficient}, temperature accelerated dynamics (\groen{TAD}) \cite{so2000temperature}, blue moon sampling \cite{den1998calculation,carter1989constrained,ciccotti2004blue}, adaptive biasing force (\groen{ABF}) \cite{darve2008adaptive}, and thermodynamic integration (\groen{TI}) \cite{kirkwood1935statistical,straatsma1988free}. \blauwr{Where ``good'' CVs coincident with the important collective motions that separate metastable states are available, CV biasing is typically more computationally efficient than tempering approaches since it performs targeted acceleration along pre-selected order parameters\cite{abrams2013enhanced,wang2017nonlinear}. In instances where such CVs are not known or poor choices are made, CV biasing can fail badly and tempering approaches generally prove more efficient and successful. It is normally difficult to intuit good CVs for all but the simplest systems.}

Machine learning presents a systematic means to discover data-driven CVs. Conceptually, these approaches posit that the molecular dynamics in the 3$N$-dimensional space of the Cartesian coordinates of the $N$ atoms are effectively restrained to low-dimensional ``intrinsic manifold'' containing the slow dynamics to which the remaining fast degrees of freedom are slaved \cite{rohrdanz2013discovering,hegger2007complex}. The validity of this assumption has been demonstrated many times and corresponds to the emergence of a small number of collective modes from cooperative couplings between system degrees of freedom \cite{ferguson2011nonlinear,hegger2007complex,ferguson2010systematic,zhuravlev2009deconstructing,amadei1993essential,garcia1992large,das2006low,stamati2010application,ichiye1991collective}. CV discovery algorithms are therefore essentially dimensionality reduction or manifold learning approaches \cite{ferguson2011nonlinear, rohrdanz2013discovering}, and the CVs are ``good'' in that they \blauwr{parameterize the intrinsic geometry of the phase space and can correspond to the slow motions of the system over the manifold} \cite{ferguson2011nonlinear,ferguson2010systematic}. Data-driven CV discovery suffers from a fundamental difficulty: to perform good sampling of configurational space one requires good CVs that are explicit functions of the atomic coordinates, but to discover these CVs one needs to perform good sampling of configurational space to generate data for CV discovery. This is the root of the biased sampling ``chicken-and-egg problem'' identified by Clementi and co-workers \cite{rohrdanz2013discovering}. Accordingly, data-driven CV discovery must typically be performed in an iterative fashion by interleaving rounds of accelerated sampling and CV discovery: each round of sampling provides more comprehensive sampling of configurational space, and each round of CV discovery determines collective variables parameterizing the slow dynamical motions over the space explored by the biased calculations. 

Linear dimensionality reduction approaches such as principal components analysis (\groen{PCA})\cite{pearson1901liii,garcia1992large,ichiye1991collective} and multidimensional scaling (\groen{MDS}) \cite{troyer1995protein} are straightforward to apply, and furnish CVs that are explicit functions of the atomic coordinates. As inherently linear approaches, however, the CVs are typically poor descriptors for complex biomolecules possessing convoluted and nonlinear intrinsic manifolds \cite{rohrdanz2013discovering,ferguson2011nonlinear,ferguson2010systematic}. Kernel PCA presents a means to alleviate this problem by applying a known nonlinear transformation of the atomic coordinates prior to dimensionality reduction\cite{scholkopf1997kernel}, but the specification of appropriate kernels can be almost as difficult as intuiting the CVs themselves \cite{ferguson2011nonlinear}. Nonlinear manifold learning approaches include local tangent space alignment \cite{wang2011geometric}, locally linear embedding (\groen{LLE}) \cite{roweis2000nonlinear,zhang2007mlle}, Isomap \cite{das2006low,tenenbaum2000global,weinberger2006unsupervised,li2006version}, Laplacian eigenmaps \cite{belkin2002laplacian}, Hessian eigenmaps \cite{donoho2003hessian}, semidefinite embedding (\groen{SDE}) / maximum variance unfolding (\groen{MVU}) \cite{weinberger2006unsupervised}, diffusion maps \cite{ferguson2011integrating,ferguson2010systematic,coifman2006diffusion,rohrdanz2011determination,preto2014fast}, and sketch maps \cite{ceriotti2011simplifying,tribello2012using,ceriotti2013demonstrating}. These approaches have demonstrated great success in parameterizing nonlinear intrinsic manifolds of complex bio- and macromolecules by discovering nonlinear CVs corresponding to concerted structural motions \cite{ferguson2011integrating,ferguson2010systematic,das2006low,stamati2010application,rohrdanz2011determination,preto2014fast,ferguson2010experimental,zheng2011polymer,zheng2013rapid}. For the purposes of CV biasing, however, these approaches all share the critical deficiency in that they do not furnish the mapping of atomic coordinates to the CVs. An explicit and differentiable mapping from the atomic coordinates to the CVs is required to perform biased MD simulations in the CVs in order to propagate biasing forces in the CVs to forces on atoms \cite{hashemian2013modeling,rohrdanz2013discovering,ferguson2011nonlinear,spiwok2011metadynamics,fiorin2013using}. (Biased Monte-Carlo requires that the CVs be explicit, but not necessarily differentiable, functions.) Without this mapping, it is not possible perform biasing directly in CVs discovered by nonlinear learning.

A small number of approaches have been developed to perform indirect and approximate biasing in CVs for which the atomic mapping is unknown. We previously introduced perhaps the simplest approach that correlates CVs with candidate physical variables (e.g., dihedral angles, radius of gyration) and performs biased sampling in these proxy variables \cite{ferguson2011integrating}. High-throughput screening can sieve putative physical variables and identify those best correlated with the CVs\cite{ma2005automatic,peters2006obtaining,peters2007extensions}. This approach is, however, typically rather unsatisfactory, since it surrenders sampling in the CVs, and is very sensitive to the quality of the physical proxies. A related approach fits the parameters of CVs with predefined functional forms \cite{abrams2012fly}, but the definition of appropriate functions can almost as difficult as defining the CVs themselves. A more sophisticated approach expresses atomic coordinates as a sum of localized basis functions over a small number of ``landmarks'' in the CV embedding \cite{hashemian2013modeling,li2006version,spiwok2011metadynamics,branduardi2007b}, and projects new data by performing a weighted nearest-neighbors embedding \cite{spiwok2011metadynamics,branduardi2007b} or solving a linear optimization problem to minimize the projection error \cite{hashemian2013modeling,li2006version}. Accordingly, direct sampling in the CVs is relinquished for an interpolative and approximate basis function expansion\cite{ferguson2011nonlinear}. An elegant strategy termed diffusion-map-directed molecular dynamics (\groen{DM-d-MD}) proposed by Clementi and co-workers accelerates sampling by initializing unbiased simulations in poorly sampled regions of the CV projection \cite{preto2014fast,zheng2013rapid}. Kevrekidis and co-workers recently proposed intrinsic map dynamics (\groen{iMapD}) as an ingenious approach that first employs diffusion maps to construct a nonlinear parameterization of the local manifold, then combines boundary detection and local principal components analysis to smoothly extend the boundary and initialize short bursts of unbiased simulations in unexplored regions of configurational space \cite{chiavazzo2017intrinsic}. Both DM-d-MD and iMapD use CVs discovered by nonlinear manifold learning to define the limits of current exploration of configurational space, then judiciously initialize new unbiased simulations to drive further exploration and accelerate sampling. However, neither approach actually performs biased sampling directly in the CVs. This can make phase space exploration inefficient by presenting difficulties for surmounting high free energy barriers down which unbiased simulations can rapidly tumble.

In this work we present a new approach based on auto-associative artificial neural networks (``autoencoders'') \cite{scholz2002nonlinear,scholz2008nonlinear,hinton2006reducing,yan2007graph,wang2014generalized} to discover nonlinear CVs and perform biasing along these coordinates. Crucially, neural networks offer a flexible and powerful means to discover nonlinear CVs and furnish a explicit and differentiable function of the atomic coordinates. These CVs may then be implemented directly in any CV biasing approach for enhanced sampling, such as umbrella sampling or metadynamics. This feature makes autoencoders uniquely suited to CV discovery for enhanced sampling, and stands in contrast to existing nonlinear machine learning approaches that require appealing to proxy variables, basis set projections, or local linear dimensionality reduction. We term our approach \underline{M}olecular \underline{E}nhanced \underline{S}ampling with \underline{A}utoencoders (\groen{MESA}). In this article we introduce the theoretical and mathematical underpinnings of MESA, and demonstrate it in applications to alanine dipeptide and Trp-cage, and make the approach freely available to the community through an open source plugin to the molecular simulation package OpenMM \cite{eastman2017openmm,eastman2012openmm,friedrichs2009accelerating}.

The structure of this paper is as follows. In the \blauw{next section}, we establish the theoretical and mathematical underpinnings of MESA including autoencoders, data-augmentation, the L-method for dimensionality identification, umbrella sampling, and the weighted histogram analysis method (\groen{WHAM}). In \blauw{Section \ref{results_discussion_section}}, we demonstrate our approach to discover nonlinear CVs, perform enhanced sampling in these coordinates, and efficiently recover free energy surfaces for alanine dipeptide and Trp-cage. In \blauw{Section \ref{sec:concl}}, we close with our conclusions and outlook for future work and development.

\section{\sffamily \Large Methods} \label{method_section}

\subsection{\sffamily \large Nonlinear dimensionality reduction using autoencoders}

\subsubsection{\sffamily \normalsize Auto-associative artificial neural networks (autoencoders)} \label{AE}

\begin{figure}[ht!]
\begin{center}
\includegraphics[width=0.95\textwidth]{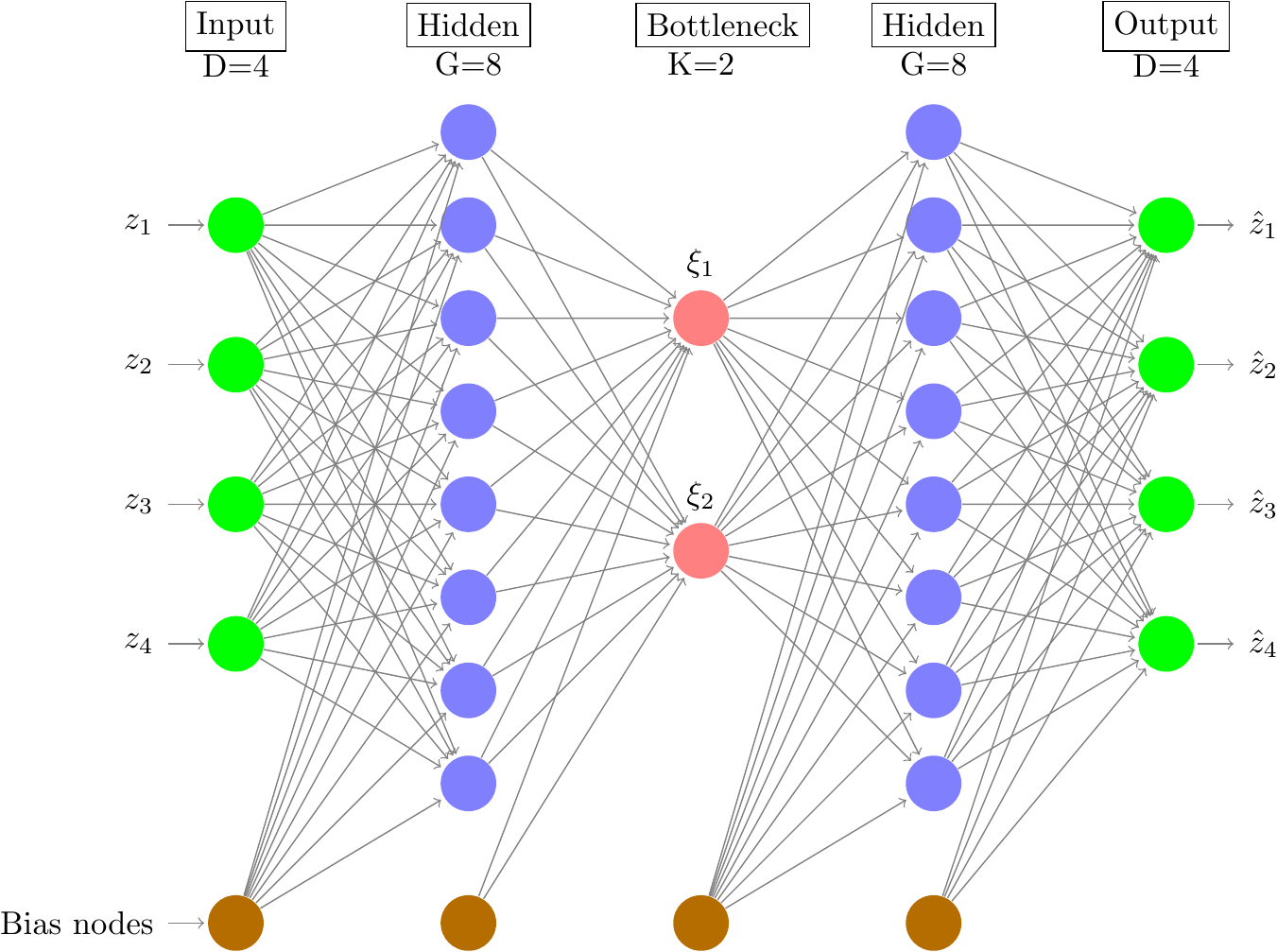}
\caption{An auto-associative artificial neural network (autoencoder) with a 4-8-2-8-4 topology. The first half of the network passes inputs $\mathbf{z}_q \in \mathbb{R}^D$ through a hidden layer into a bottleneck layer containing the low-dimensional nonlinear projection $\boldsymbol\xi_q \in \mathbb{R}^K$. The second half of the network passes the $\boldsymbol\xi_q \in \mathbb{R}^K$ through a hidden layer to approximately reconstruct the inputs in the output layer $\hat{\mathbf{z}}_q \in \mathbb{R}^D$. Each node sums input signals from nodes in the previous layer according to adjustable weights $\mathbf{w}$ plus a constant bias $\mathbf{b}$ and passes them through an activation function to generate its own output. Training the network parameters $\{\mathbf{w},\mathbf{b}\}$ to accurately reproduce its inputs (i.e., to autoencode) corresponds to discovery of a low-dimensional projection in the bottleneck layer capturing the salient features of the data. Image constructed using code downloaded from \url{http://www.texample.net/tikz/examples/neural-network} with the permission of the author Kjell Magne Fauske.} \label{autoencoder}
\end{center}
\end{figure}

Artificial neural networks (\groen{ANN}) are a collection of simple computational nodes (``artificial neurons'') linked together into a network\cite{hassoun1995fundamentals}, and so-called because of their similarity to biological neural networks. ANNs have risen to prominence as a flexible tool to perform nonlinear regression and classification due to their large expressive power and predictive capacity \cite{mcculloch1943logical}, with simple multilayer networks proven to be universal approximators of any nonlinear input-output relationship\cite{hassoun1995fundamentals,chen1995universal}. In this work, we focus on a class of auto-associative ANNs known as autoencoders, which are specifically designed to perform unsupervised nonlinear dimensionality reduction and furnish explicit and differentiable functions for the nonlinear projection\cite{scholz2008nonlinear,kramer1991nonlinear,kirby1996circular}. The structure of a prototypical 5-layer autoencoder is shown in \blauw{Fig.~\ref{autoencoder}}. \blauwr{In this section we review the key mathematical details of these networks that we exploit within our CV discovery and enhanced sampling framework.} 

\textbf{Architecture.} The fundamental ansatz of the autoencoder is that there exists a low-dimensional subspace -- the so-called ``intrinsic manifold'' \cite{ferguson2010systematic} -- $\mathcal{L} \in \mathbb{R}^K$ containing most of the variance in the high-dimensional data $\mathcal{H} \in \mathbb{R}^D$. Training of the autoencoder can discover both the nonlinear projection of the high-dimensional data to the intrinsic manifold $\Theta_\mathrm{proj} : \mathcal{H} \rightarrow \mathcal{L}$ and an approximate reconstruction of the high-dimensional states from their low-dimensional projections $\Theta_\mathrm{rec} : \mathcal{L} \rightarrow \mathcal{H}$\cite{scholz2008nonlinear,bourlard1988auto}. The architecture of the network encodes this ansatz within a five-layer, symmetric, fully-connected, feedforward topology. The projection function $\Theta_\mathrm{proj}$ is encoded in the first three layers, in which an ensemble of $Q$ high-dimensional data points $\{\mathbf{z}_q\}_{q=1}^Q \in \mathbb{R}^D$ comprising system snapshots harvested from a molecular dynamics trajectory, are fed to an \textit{input layer} containing $D$ nodes, passed to the \textit{first hidden layer} containing $G$ nodes, and finally projected down into a \textit{bottleneck layer} comprising $K$ nodes containing the low-dimensional projections $\{\boldsymbol\xi_q\}_{q=1}^Q \in \mathbb{R}^K$. The reconstruction function $\Theta_\mathrm{rec}$ is encoded in the last three layers, in which the low-dimensional projections $\{\boldsymbol\xi_q\}_{q=1}^Q \in \mathbb{R}^K$ in the bottleneck layer are passed through the \textit{second hidden layer} of $G$ nodes, and finally to the $D$-node \textit{output layer} containing the reconstructed data points $\{\hat{\mathbf{z}}_q\}_{q=1}^Q \in \mathbb{R}^D$. This is known as a $D$-$G$-$K$-$G$-$D$ autoencoder architecture. The number of input and output nodes $D$ is specified by the dimensionality of the data. The number of hidden nodes is flexible and may be tuned via cross-validation to trade-off network expressive flexibility against complexity. Choosing $G \approx 2D$ often provides a good initial guess. The number of bottleneck nodes $K$ specifies the intrinsic dimensionality of the system. We present below an automated approach to ascertain an appropriate value of $K$.

\textbf{Activation functions.} Denoting the \textit{input}, \textit{output}, and \textit{bias} associated with node $j$ of layer $i$ as $x_j^{(i)}$, $y_j^{(i)}$, and $b_j^{(i)}$, the \textit{weight} of the connection between node $j$ in layer $i$ and node $k$ in layer $(i+1)$ as $w_{jk}^{(i)}$, and the \textit{activation function} of layer $i$ as $f^{(i)}$, we can build up an explicit mathematical mapping for $\mathbf{z}_q \rightarrow \boldsymbol\xi_q \rightarrow \hat{\mathbf{z}}_q$. The inputs to the input layer are simply the components of the high dimensional data points $x_j^{(1)} = z_j$. The input to any node in any other layer is the weighted sum of the outputs of all of the nodes in the previous layer plus a bias $x_k^{(i)} = b_k^{(i)} + \sum_j w_{jk}^{(i-1)} y_j^{(i-1)}$, and the output of that node is the result of activation function acting on its input $y_k^{(i)} = f^{(i)}\left( x_k^{(i)} \right)$. The input layer possesses linear activation functions (i.e., identity mappings) whereas the remaining layers possess nonlinear activation functions that we select to be hyperbolic tangents\cite{friedman2001elements}. Importantly for our application, these smooth activation functions possess analytical first derivatives. Summing over the first three layers, the analytical expression for the nonlinear projection is,
\begin{equation}\label{proj}
\Theta_\mathrm{proj} : \xi_k = f^{(3)} \left( b_k^{(3)} + \sum_{g=1}^G w_{gk}^{(2)} f^{(2)} \left( b_g^{(2)} + \sum_{d=1}^D w_{dg}^{(1)} z_d \right) \right),
\end{equation}
and for the nonlinear reconstruction is,
\begin{equation}\label{rec}
\Theta_\mathrm{rec} : \hat{z}_d = f^{(5)}\p{b_d^{(5)} + \sum_{g=1}^G w_{gd}^{(4)} f^{(4)} \left( b_g^{(4)} + \sum_{k=1}^K w_{kg}^{(3)} \xi_k \right)}.
\end{equation}

\textbf{Training.} Training of the autoencoder amounts to training the network to optimally reconstruct its own inputs (i.e., to autoencode). Mathematically, this corresponds to specifying the network weights $\mathbf{w}$ and biases $\mathbf{b}$ to minimize the error function,
\begin{align}\label{error_fun}
E &= \sum_{q=1}^Q \left\Vert \mathbf{z}_q - \hat{\mathbf{z}}_q \right\Vert^2 + \sum_{i,j,k} \lambda_i \left( w_{jk}^{(i)} \right)^2 = \sum_{q=1}^Q \left\Vert \mathbf{z}_q - (\Theta_\mathrm{rec} \circ \Theta_\mathrm{proj}) \mathbf{z}_q \right\Vert^2 + \sum_{i,j,k} \lambda_i \left( w_{jk}^{(i)} \right)^2.
\end{align}
\blauwr{The first term is the reconstruction error, and the second term is a regularizing penalty known as \textit{weight decay} that is typically included to prevent overfitting by controlling the growth of the network weights \cite{scholz2008nonlinear,friedman2001elements}. In this work, we choose to set $\lambda_i = 0$ for all layers to eliminate the weight decay term and instead regularize network training using early stopping as an approach that plays essentially the same role in guarding against overfitting without the need to tune the $\lambda_i$ hyperparameters \cite{collobert2004links,bengio2012practical}. Early stopping is implemented by randomly selecting 80\% of data as the training partition and remaining 20\% as the validation partition, and terminating training when we either reach the maximum number of training epochs or the error function evaluated over the validation partition no longer decreases for 30 continuous epochs (i.e., early stopping). In each epoch, neural network parameters are updated using mini-batch stochastic gradient descent with momentum. \cite{hassoun1995fundamentals,rumelhart1995backpropagation,rumelhart1988learning,sutskever2013importance}.}

\textbf{Dimensionality determination.} The number of nodes in the bottleneck layer $K$ specifies the dimensionality of the nonlinear projection. We empirically determine the dimensionality of the intrinsic manifold from the data by training $D$-$G$-$K$-$G$-$D$ autoencoder architectures, where $G$ is tuned via cross-validation \cite{friedman2001elements}. An appropriate value of $K$ is determined by computing the fraction of variance explained (\groen{FVE}) by the reconstructed outputs,
\begin{align}\label{FVE}
\mathrm{FVE} &= 1 - \frac{\mathrm{SS_{err}}}{\mathrm{SS_{tot}}} = 1 - \frac{\sum_{q=1}^Q \left(\mathbf{z}_q - \hat{\mathbf{z}}_q \right)^2}{\sum_{q=1}^Q \left(\mathbf{z}_q - \bar{\mathbf{z}}_q \right)^2}
\end{align}
where $\mathrm{SS_{err}}$ is the residual sum of squares, $\mathrm{SS_{tot}}$ is the total sum of squares, and $\bar{\mathbf{z}}$ is the mean input vector. The emergence of a plateau or knee in the FVE curve at a particular value of $K$ indicates that the predominance of variance in the data exists within a $K$-dimensional intrinsic manifold. We determine the location of the knee in an automated fashion using the L-method of Salvador and Chan \cite{salvador2004determining}. For $K \ll D$ and a high FVE in excess of 70\%, we assert that the autoencoder has discovered that the high-dimensional data in $\mathbb{R}^D$ admits a low-dimensional representation in $\mathbb{R}^K$ from which the original data can be approximately reconstructed. The outputs of the bottleneck layer $\boldsymbol\xi$ in the trained autoencoder define the CVs parameterizing the low-dimensional intrinsic manifold. The explicit mapping of the high-dimensional input data to these CVs is furnished by the projection $\Theta_\mathrm{proj}$ in \blauw{Eqn.~\ref{proj}}. Provided the activation functions of all of the nodes in the network are smooth functions, then this explicit mapping also possesses analytical first derivatives with respect to the input data.

\subsubsection{\sffamily \normalsize Elimination of translational and rotational invariances} \label{dataAug}

\begin{figure}[ht!]
\begin{center}
\includegraphics[width=0.95\textwidth]{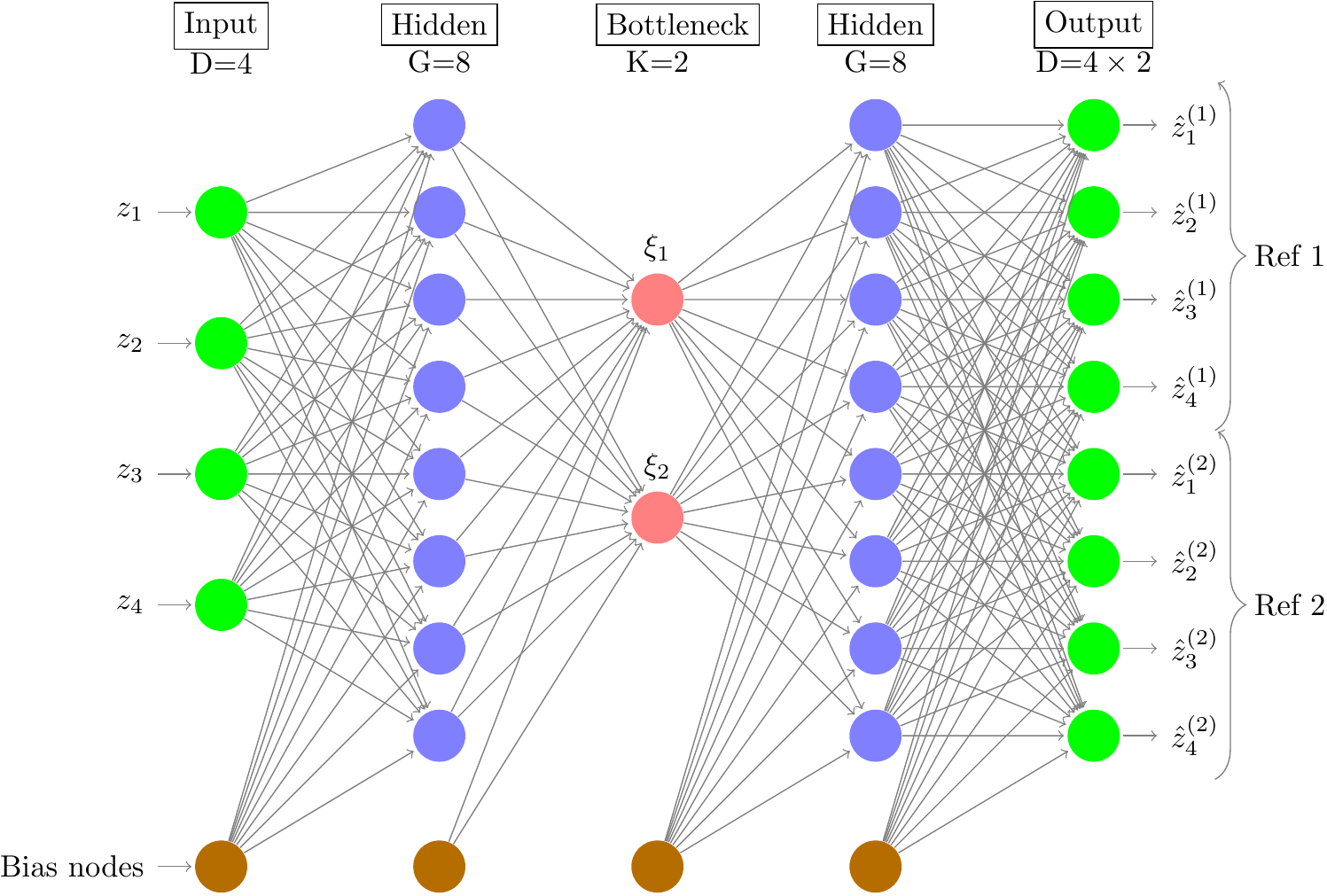}
\caption{An autoencoder with a 4-8-2-8-(2$\times$4) topology). The network topology is identical to that in \blauw{Fig.~\ref{autoencoder}}, with the exception that the first four outputs of the output layer are used to compute the reconstruction error with respect to the first reference configuration, and the second four outputs with respect to the second reference configuration. The network is trained to minimize the sum of reconstruction errors over the two reference configurations. Image constructed using code downloaded from \url{http://www.texample.net/tikz/examples/neural-network} with the permission of the author Kjell Magne Fauske.} \label{autoencoder_ref_2}
\end{center}
\end{figure}

Our application of autoencoders is to learn nonlinear low-dimensional CVs describing the important configurational motions of biomolecules. The training data $\{\mathbf{z}_q\}_{q=1}^Q$ comprises $Q$ snapshots of the molecule harvested from molecular dynamics simulations recording the Cartesian coordinates of the constituent atoms. When studying the configurational motions of a molecule in an isotropic medium, we typically wish to resolve internal structural reconfigurations and exclude trivial changes in rotational orientation or center of mass location.  Accordingly, we must properly treat the $\mathbf{z}_q$ to eliminate rotation and translation. It is straightforward to eliminate translational degrees of freedom by subtracting the center of mass coordinates from each $\mathbf{z}_q$ prior to passing these data to the autoencoder. 
We eliminate rotational invariances using data augmentation approach that is commonly used in deep learning to eliminate invariances and enhance training sets, and which has proved particularly valuable in image classification applications \cite{krizhevsky2012imagenet}. For each translationally-centered molecular configuration $\mathbf{z}_q$, we generate an ensemble of $N$ additional configurations constructed by applying random three-dimensional rotations to $\mathbf{z}_q$. In training the autoencoder we then assert that the network learn that all of these inputs should ``autoencode'' to the same output, thereby teaching the network to disregard rotations as a relevant degree of freedom and discover CVs that are a function only of the internal molecular configuration. Mathematically, this requires that we modify the error function defined in \blauw{Eqn.~\ref{error_fun}} to,
\begin{align}\label{err_2}
E &= \sum_{q=1}^Q \sum_{n=1}^N \left\Vert \hat{\mathbf{z}}_q - L \left( R_n \left( \mathbf{z}_q \right) ,  \mathbf{z}_\mathrm{ref} \right) \right\Vert^2 + \sum_{i,j,k} \lambda_i \left( w_{jk}^{(i)} \right)^2,
\end{align}
\blauwr{where $R_n$ denotes a rotation of configuration $\mathbf{z}_q$ by a randomly selected angle $\theta_n = [0,2\pi)$ around a randomly selected axis $\mathbf{u}_n$ in 3D space, and $L \left( R_n \left( \mathbf{z}_q \right),  \mathbf{z}_\mathrm{ref} \right)$ represents the optimal rotational alignment of rotated configuration $R_n \left( \mathbf{z}_q \right)$ to a reference configuration $\mathbf{z}_\mathrm{ref}$ that has been previously translated such that its center of mass is coincident with the origin. The operation $L \left( R_n \left( \mathbf{z}_q \right),  \mathbf{z}_\mathrm{ref} \right)$ can be efficiently performed using the Kabsch algorithm \cite{kabsch1976solution}.} Training of the autoencoder weights and biases to minimize \blauw{Eqn.~\ref{err_2}} teaches the network to map arbitrary rotations of a particular configuration to the alignment of that configuration to a specific reference structure. A deficiency of this approach is that we must select a particular reference configuration $\mathbf{z}_\mathrm{ref}$ and this selection can affect the particular set of trained weights and biases. We observe that bias introduced by the choice of reference state can be mitigated and the autoencoder made robust to this particular choice by employing $T$ reference configurations $\{ \mathbf{z}_\mathrm{ref}^t \}_{t=1}^T$, each of which is reconstructed by a subset of the outputs of an augmented terminal layer. For example, we show in \blauw{Fig.~\ref{autoencoder_ref_2}} an autoencoder with 4-8-2-8-(2$\times$4) topology employing two reference configurations. The network is identical to that in \blauw{Fig.~\ref{autoencoder}}, except that approximations to each of the two reference configurations are computed in the terminal layer. We then define the error function as the sum of the errors associated with each reference structure,
\begin{align}\label{err_3}
E &= \sum_{t=1}^T \sum_{q=1}^Q \sum_{n=1}^N \left\Vert \hat{\mathbf{z}}_q^t - L \left( R_n \left( \mathbf{z}_q \right) ,  \mathbf{z}_\mathrm{ref}^t \right) \right\Vert^2 + \sum_{i,j,k} \lambda_i \left( w_{jk}^{(i)} \right)^2,
\end{align}
allowing the weights and biases of the output layer to differ for each of the $T$ reference configurations. In this manner, we confine the effect of different reference structure rotations to the parameters of the terminal layer, and the weights and biases of the first four layers are trained to be invariant to the choice of reference structure. In practice, we find satisfactory performance for autoencoders trained with $T$=2 reference structures and $N$=16 random rotations.

\subsection{\sffamily \large Enhanced sampling in autoencoder CVs}

\subsubsection{\sffamily \normalsize Umbrella sampling}

The principal advantage of autoencoders over competing nonlinear dimensionality reduction techniques is that the mapping of the high dimensional data $\{\mathbf{z}_q\}_{q=1}^Q \in \mathbb{R}^D$ -- here molecular configurations -- to the intrinsic manifold $\{\boldsymbol\xi_q\}_{q=1}^Q \in \mathbb{R}^K$ spanned by the discovered CVs is an explicit and differentiable function of the input coordinates given by \blauw{Eqn.~\ref{proj}}. Accordingly, biasing forces on the CVs can be analytically propagated to biasing forces on the atoms by repeated application of the chain rule. Enhanced sampling can then be conducted directly in the discovered CVs using any of the collective variable biasing techniques listed in \blauw{Section \ref{sec:intro}}. In this work we choose to use umbrella sampling, but other popular approaches such as metadynamics or adaptive biasing force calculations could be straightforwardly employed.

Umbrella sampling (\groen{US}) was first introduced by Torrie and Valleau in 1977 \cite{torrie1977nonphysical}. The method proceeds by augmenting the configurational part of the classical Hamiltonian $H(\mathbf{r}^N)$ written in the coordinates of all constituent atoms $\mathbf{r}^N$ with a (typically harmonic) restraining potential $V(\boldsymbol\xi(\mathbf{r}^N))$ in a set of predefined collective variables $\boldsymbol\xi(\mathbf{r}^N) \in \mathbb{R}^K$. These potentials confine the system to a narrow region of CV space that looks locally flat, and within which the system can comprehensively sample the acessible configurational space\cite{roux1995calculation, chipot2007free}. By tiling CV space with a number of overlapping umbrella windows, free energy barriers can be surmounted and the full extent of CV space completely explored. Success of the approach depends crucially on the selection of ``good'' CVs that track the slow collective molecular motions of the system and adequately discriminate the metastable states and resolve the free energy barriers within the system.

The bias-augmented system Hamiltonian $E(\mathbf{r}^N) = H(\mathbf{r}^N) + V(\boldsymbol\xi(\mathbf{r}^N))$ determines the force $\mathbf{f}_i$ experienced by each atom $i$ through the first derivatives with respect to the atomic position $\mathbf{r}_i$,
\begin{align}
\mathbf{f}_i &= -\nabla_{\mathbf{r}_i} E(\mathbf{r}^N) \notag \\
&= -\nabla_{\mathbf{r}_i} H(\mathbf{r}^N) - \nabla_{\mathbf{r}_i} V(\boldsymbol\xi(\mathbf{r}^N)) \notag \\
&= \mathbf{f}_i^\mathrm{U} + \mathbf{f}_i^\mathrm{B},
\end{align}
where we identify $\mathbf{f}_i^\mathrm{U} = -\nabla_{\mathbf{r}_i} H(\mathbf{r}^N)$ as the unbiased forces on the atoms arising from the system Hamiltonian in the absence of the biasing potential, and $\mathbf{f}_i^\mathrm{B} = - \nabla_{\mathbf{r}_i} V(\boldsymbol\xi(\mathbf{r}^N))$ as the artificial biasing forces on the atoms arising from the potential applied to the CVs. In performing a biased molecular dynamics simulation, the unbiased forces are computed from the molecular force field in exactly the same manner as in an unbiased calculation. These are supplemented by the biased forces derived from the artificial restraining potential applied to the CVs. We now proceed to derive analytical expressions for the biasing forces. In \blauw{Section \ref{implementation}} we describe our computational implementation within biased molecular dynamics simulations.

We first recognize that the $\boldsymbol\xi \in \mathbb{R}^K$ furnished by the bottleneck layer depend on the inputs to the autoencoder $\mathbf{z} \in \mathbb{R}^D$ through the map defined in \blauw{Eqn.~\ref{proj}}. In applications of nonlinear machine learning to biomolecular folding it is conventional to treat the solvent atom coordinates implicitly through their influence on the conformational ensemble explored by the biomolecule, but it is a possible extension of this work to explicitly consider solvent degrees of freedom \cite{ferguson2011nonlinear}. Accordingly, the autoencoder inputs $\mathbf{z}$ comprise a list of the Cartesian coordinates of the $D/3$ atoms constituting the biomolecule $\mathbf{z} = \{z_1^x, z_1^y, z_1^z, \ldots, z_{D/3}^x, z_{D/3}^y, z_{D/3}^z,\}$ as a subset of the coordinates of all atoms in the system $\mathbf{z} \subset \mathbf{r}^N$. Armed with these nested dependencies, we may expand $\mathbf{f}_i^\mathrm{B}$ as a product of three Jacobian matrices resulting from repeated application of the chain rule,
\begin{align} \label{fiB}
\mathbf{f}_i^\mathrm{B} &= - \frac{\partial V}{\partial \boldsymbol\xi} \frac{\partial \boldsymbol\xi}{\partial \mathbf{z}} \nabla_{\mathbf{r}_i} \mathbf{z} \notag \\
&= \left[ -\sum_{k=1}^K \sum_{d=1}^D \frac{\partial V}{\partial \xi_k} \frac{\partial \xi_k}{\partial z_d} \frac{\partial z_d}{\partial r_i^x} \quad -\sum_{k=1}^K \sum_{d=1}^D \frac{\partial V}{\partial \xi_k} \frac{\partial \xi_k}{\partial z_d} \frac{\partial z_d}{\partial r_i^y} \quad -\sum_{k=1}^K \sum_{d=1}^D \frac{\partial V}{\partial \xi_k} \frac{\partial \xi_k}{\partial z_d} \frac{\partial z_d}{\partial r_i^z}\right]
\end{align}
where $\frac{\partial V}{\partial \boldsymbol\xi}$ is a 1-by-$K$ matrix, $\frac{\partial \boldsymbol\xi}{\partial \mathbf{z}}$ is a $K$-by-$D$ matrix, and $\nabla_{\mathbf{r}_i} \mathbf{z}$ is a $D$-by-3 matrix \cite{hashemian2013modeling}. Each of these three matrices possesses straightforward analytical expressions.

\fbox{$\frac{\partial V}{\partial \boldsymbol\xi}$} Assuming the standard choice of a harmonic biasing potential in each of the CVs, the artificial biasing potential can be written as,
\begin{align} \label{potential_fun}
V(\boldsymbol\xi) = \sum_{k=1}^K \frac{1}{2} \kappa_k \left( \xi_k - \xi_k^0 \right)^2,
\end{align}
where $\kappa_k$ is the harmonic force constant and $\xi_k^0$ the harmonic center for CV $\xi_k$. The elements of the Jacobian matrix immediately follow as,
\begin{align} \label{dVdxi}
\frac{\partial V}{\partial \boldsymbol\xi_k} = \kappa_k \left( \xi_k - \xi_k^0 \right).
\end{align}

\fbox{$\frac{\partial \boldsymbol\xi}{\partial \mathbf{z}}$} Appealing to \blauw{Eqn.~\ref{proj}}, the elements of this Jacobian matrix follow from the low-dimensional projection furnished by the trained autoencoder,
\begin{align} \label{dxidz}
\frac{\partial \xi_k}{\partial z_d} &= \frac{\partial}{\partial z_d} \left[ f^{(3)} \left( b_k^{(3)} + \sum_{g=1}^G w_{gk}^{(2)} f^{(2)} \left( b_g^{(2)} + \sum_{d=1}^D w_{dg}^{(1)} z_d \right) \right) \right] \notag \\
&= f^{(3)\prime} \left( b_k^{(3)} + \sum_{g=1}^G w_{gk}^{(2)} f^{(2)} \left( b_g^{(2)} + \sum_{d=1}^D w_{dg}^{(1)} z_d \right) \right) \times \sum_{g=1}^G w_{gk}^{(2)} f^{(2)\prime} \left( b_g^{(2)} + \sum_{d=1}^D w_{dg}^{(1)} z_d \right) \times w_{dg}^{(1)}
\end{align}
where the $\mathbf{w}$ and $\mathbf{b}$ are the weights an biases of the trained network, $f^{(2)}$ and $f^{(3)}$ are the $tanh$ activation functions associated with the second and third layers, and $f^{(2)\prime}$ and $f^{(3)\prime}$ are their first derivatives $tanh(x)^\prime = 1 - tanh^2(x)$. 

\fbox{$\nabla_{\mathbf{r}_i} \mathbf{z}$} Since the $\mathbf{z}$ constitute a subset of the atomic Cartesian coordinates, this final Jacobian matrix possesses a very simple form. For $\mathbf{r}_i \subset \mathbf{z}$ the $D$-by-3 matrix contains three unit elements $\frac{\partial z_d}{\partial r_i^x} = 1$ for $z_d = r_i^x$, $\frac{\partial z_d}{\partial r_i^y} = 1$ for $z_d = r_i^y$, and $\frac{\partial z_d}{\partial r_i^z} = 1$ for $z_d = r_i^z$, with all other entries zero. For $\mathbf{r}_i \not\subset \mathbf{z}$, all matrix elements are zero. Accordingly, this matrix serves as a filter that passes forces onto only those atoms whose coordinates feature in the inputs to the autoencoder.

\subsubsection{\sffamily \normalsize Weighted histogram analysis method (WHAM)} \label{WHAM}

Having conducted a series of umbrella sampling simulations tiling CV space, we estimate the unbiased free energy surface $F(\boldsymbol\xi)$ in these CVs by reweighting and combining the biased histograms recorded in each window using the weighted histogram analysis method (\groen{WHAM}) \cite{ferguson2017bayeswham,kumar1992weighted,bartels2000analyzing,ferrenberg1989optimized,bennett1976efficient}. The WHAM estimator minimizes statistical errors in the estimated unbiased probability distribution $P(\boldsymbol\xi)$ \cite{kumar1992weighted,bartels2000analyzing,ferrenberg1989optimized,bennett1976efficient,ferguson2017bayeswham,bartels1997multidimensional,habeck2012bayesian,zhu2012convergence,gallicchio2005temperature}, and the 
free energy surface is then computed from the estimated unbiased distribution using the standard statistical mechanical relation\cite{hartmann2011two},
\begin{align} \label{unbiasedFE}
F(\boldsymbol\xi) &= -k_B T \ln{P(\boldsymbol\xi)} + C,
\end{align}
where $C$ is an arbitrary additive constant reflecting our ignorance of the absolute free energy scale.
Free energy surfaces in (collective) variables other than those in which biased sampling was conducted are estimated from the biased simulation data and WHAM solution using the approach detailed in Ref.~[\cite{ferguson2011integrating}].

\subsection{\sffamily \large MESA: Interleaved on-the-fly CV discovery and enhanced sampling}\label{integrate_US_with_autoencoder}

\begin{figure}[ht!]
\begin{center}
\includegraphics[width=0.5\textwidth]{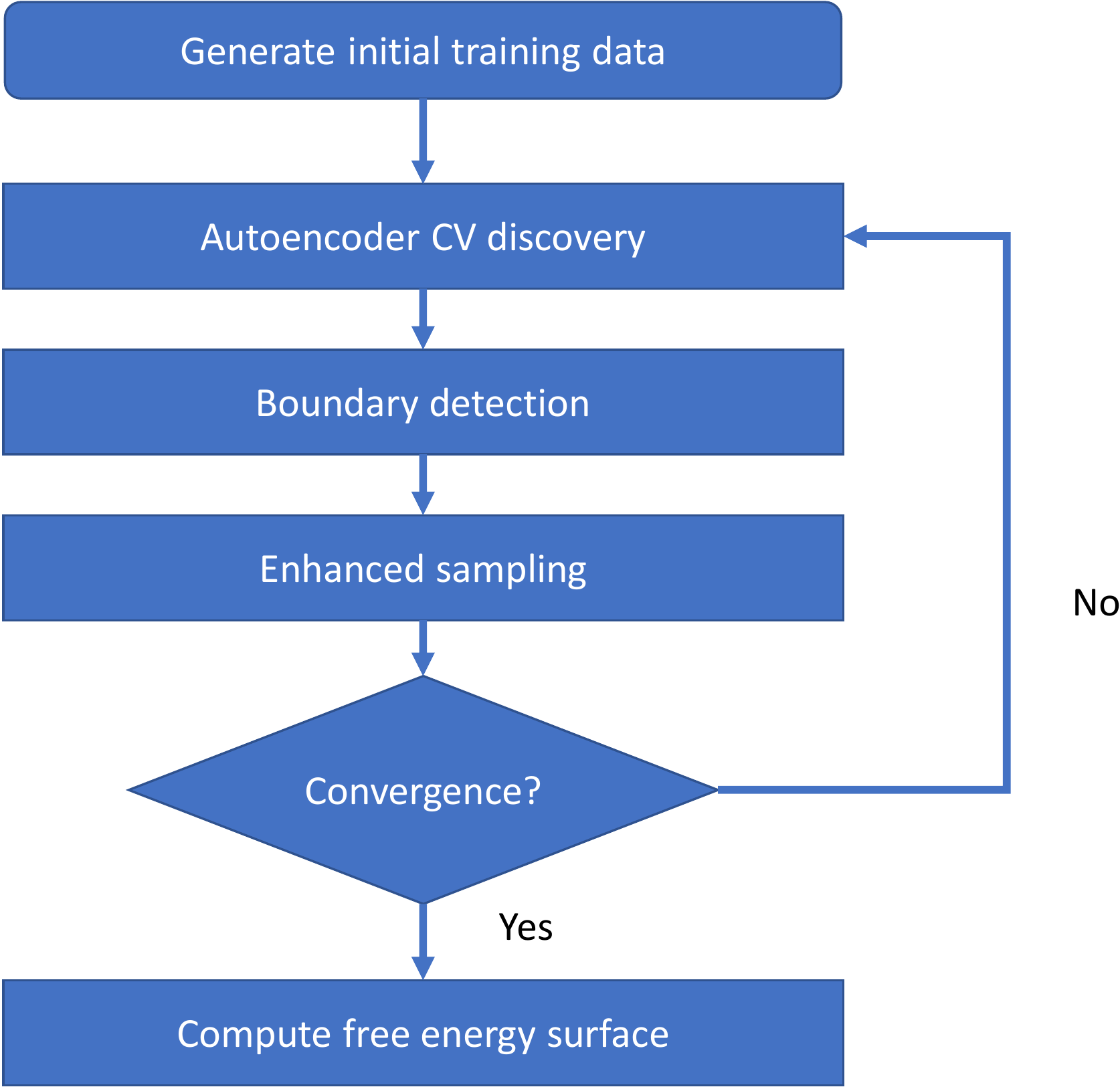} 
\caption{Schematic flowchart of the iterative six-step application of MESA.} \label{flowchart}
\end{center}
\end{figure}

\blauwr{We now present our novel framework for nonlinear CV discovery and accelerated sampling that we term MESA.} The heart of the approach is discovery of CVs that are explicit and differentiable functions of the atomic coordinates that can then be used to perform enhanced sampling by surgical targeting of biased calculations in these collective coordinates. Having conducted additional sampling of the space, the data driven CVs spanning this expanded space will, in general, differ from those parameterizing the original space and we must retrain our autoencoder including the new data. Furthermore, the dimensionality of the system may change with location in configurational space, and new CVs can emerge as we conduct additional sampling \cite{ferguson2011integrating,chiavazzo2017intrinsic}. This is the origin of the ``chicken-and-egg problem'' identified by Clementi and co-workers \cite{rohrdanz2013discovering}, and requires that we interleave successive rounds of CV discovery and biased sampling until the discovered CVs and the free energy surface they parameterize stabilize. This informs a six-step iterative protocol for the application of MESA, for which a schematic flowchart is provided in \blauw{Fig.~\ref{flowchart}}. \blauwr{We observe that the computational cost associated with autoencoder training is vastly lower than that of the biased molecular simulations for all but the most trivial of molecular systems. For example, execution of 15 $\times$ 2 ns biased simulations of the 20-residue Trp-cage mini-protein \cite{neidigh2002designing} in a bath of 2773 explicit water molecules required $\sim$8 GPU-hours, whereas training of an autoencoder over the 32,000 harvested molecular configurations required only $\sim$5 GPU-minutes.}

\vspace{0.25in}

\textbf{Step 1 -- Generate initial training data.} Perform an initial unbiased simulation to generate an initial set of atomic coordinates for the first round of autoencoder training. The CV discovery process performs better for larger data sets that explore larger volumes of configurational space, and so it is typically beneficial to conduct relatively long unbiased calculations. As a rule of thumb, the initial unbiased run is sufficiently good for initial CV discovery if the fraction of variance explained by the reconstructed outputs exceeds $\sim$40\% (\blauw{Eqn.~\ref{FVE}}). For systems possessing very rugged free energy surfaces, it can be useful to conduct this unbiased run at elevated temperature, employ a biased calculation in intuited CVs based on expert knowledge, or run calculations of a related but less computationally expensive model such as a biomolecule in vacuum or implicit solvent or a coarse-grained analog as a form of transfer learning\cite{hashemian2013modeling, sultan2017transfer}.

\textbf{Step 2 -- Autoencoder CV discovery.} Train an ensemble of $D$-$G$-$K$-$G$-$D$ autoencoders with different values of $K$ corresponding to various numbers of bottleneck nodes. $G$ is tuned at each value of $K$ by cross-validation. Perform training over all molecular configurations collected in all previous rounds of biased and unbiased sampling using the data-augmented training procedure described in \blauw{Section \ref{dataAug}} in which we eliminate translational and rotational invariances. It is typically good practice to train a number of autoencoders at each value of $K$ by initializing each with different randomly assigned weights and biases and selecting the top performed measured in terms of fraction of variance explained in the reconstructed outputs (\blauw{Eqn.~\ref{FVE}}). Apply the L-method\cite{salvador2004determining} to identify a knee in the FVE plot to determine the dimensionality of the intrinsic manifold and select an appropriate value of $K$.

\textbf{Step 3 -- Boundary detection.} Map all molecular configurations $\{\mathbf{z}_q\}_{q=1}^Q \in \mathbb{R}^D$ to the low-dimensional intrinsic manifold $\{\boldsymbol\xi_q\}_{q=1}^Q \in \mathbb{R}^K$ using the trained autoencoder. Employ a boundary detection algorithm to define the $(K-1)$-dimensional boundary of the explored region of the $K$-dimensional manifold. Possible approaches include alpha-shapes \cite{chiavazzo2017intrinsic,edelsbrunner1983shape,edelsbrunner1994three} or ``wrapping'' algorithms\cite{edelsbrunner2003surface}. Here we employ a grid-based approach motivated by the normalized graph Laplacian that discretizes CV space to identify cells that are both sparsely populated and adjacent to densely populated regions.

\begin{enumerate}

\item Divide the space spanned by the collected data into a cubic grid comprising cells with side lengths $\{l_k\}_{k=1}^K$. The size of the cells control how finely we resolve the boundary and how far we extrapolate beyond the explored region. The optimal values are system dependent, but we find the algorithm to be quite robust to the particular choice. As a rule of thumb, we find taking $l_k$ equal to $\sim$$1/20$ of the linear range of the exploration in $\xi_k$ provides satisfactory results. Collect histograms over this grid and let $n_i$ denote the number of points falling within cell $i$.

\item Compute $p_i=-e^{-n_i}$ for each cell. This operation applies a Laplacian / exponential kernel with unit variance to each cell count, which has the effect of amplifying difference between empty cells and non-empty cells. 

\item Let $K_i$ be the set containing indexes of neighboring cells of cell $i$. Compute $v_i = p_i-\frac{1}{\lvert K_i \rvert}\sum_{j \in K_i} p_j$. This operation measures the difference in the population of each grid cell with the mean population of its neighbors.

\item Discard all cell indexes with positive $v_i$ values. Sort the remaining $v_i$ and select the $\nu$ \blauwr{most negative} values (i.e., largest absolute values) as the cells in which to launch biased umbrella sampling runs. In this work we select $\nu$=10-20, but this can be tuned based on available computational resources.

\end{enumerate}

\begin{figure}[ht!]
\begin{center}
\includegraphics[width=0.8\textwidth]{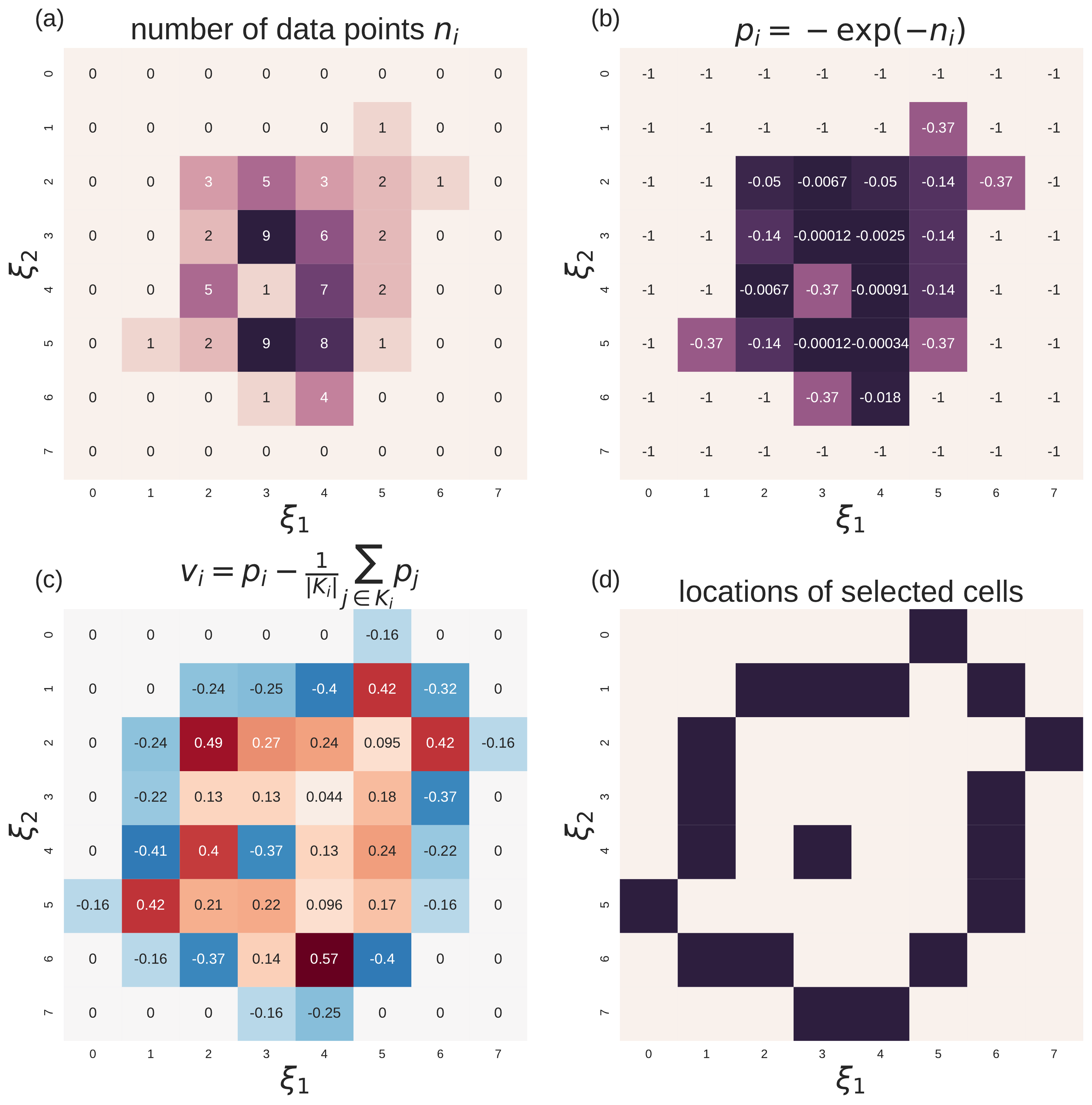} 
\caption{Illustration of boundary detection procedure for a two-dimensional intrinsic manifold parameterized by CVs $\xi_1$ and $\xi_2$. (a) Bin counts $n_i$ of molecular configurations projected into each cell of a cubic histogram spanning the explored region of the intrinsic manifold. (b) Application of Laplacian kernel to each bin count, $p_i = -e^{-n_i}$. (c) Differencing the Laplacian kernel of each cell count with the mean of its neighboring cells $K_i$, $v_i = p_i-\frac{1}{\lvert K_i \rvert}\sum_{j \in K_i} p_j$. (d) Selection of those cells containing the $\nu$ = 19 \blauwr{most negative} $v_i$ values as those in which to launch umbrella sampling runs.} \label{diagram_of_finding_boundary}
\end{center}
\end{figure}

We find this procedure to perform quite well in both identifying the boundary of the explored region of the intrinsic manifold, and identifying internal holes within this region. It is also quite robust to the shape and dimensionality of the explored region. An illustration of the boundary detection procedure is illustrated in \blauw{Fig.~\ref{diagram_of_finding_boundary}}.
 
\textbf{Step 4 -- Enhanced sampling.} Run umbrella sampling runs in each of the $\nu$ identified boundary cells employing harmonic restraining potentials given by \blauw{Eqn.~\ref{potential_fun}}. This step both advances the frontier of the explored region and fills internal holes by direct biasing in the identified CVs. In this work we select the umbrella potential center $\boldsymbol\xi^0$ to be the center of the identified cubic cells, but it is straightforward to place the biased runs closer or further from the boundary. Appropriate values for the harmonic restraining potentials $\boldsymbol\kappa$ are strongly system dependent, and must typically be tuned by trial and error. For systems possessing high free energy barriers, we suggest that it may be useful to adaptively tune $\boldsymbol\kappa$ based on the distance between the center of the umbrella sampling point cloud and the potential center. Specifically, if the center of mass of the point generated by the biased calculation falls outside of the grid cell, the harmonic force constant may be progressively increased up to a pre-specified maximum. In this manner, the biased simulation can adapt to the local gradient and topography of the underlying free energy landscape \cite{chiavazzo2017intrinsic}.

\textbf{Step 5 -- Convergence assessment.} Determine whether the umbrella sampling runs have led to substantial new exploration of configurational space. We test this in two different ways. First, we project the new molecular configurations onto the intrinsic manifold using the autoencoder and determine whether the boundary of the explored region has changed after the addition of the new data. Second, we perform \blauw{Step 2} to retrain autoencoders on the augmented data and identify whether the dimensionality of the intrinsic manifold and its parameterizing CVs have converged. The dimensionality of the intrinsic manifold is assessed by application of the the L-method\cite{salvador2004determining} to identify a knee in the FVE plot. Stabilization of the CVs is assessed by performing a linear regression of the old CVs in terms of the new, and looking for coefficients of determination in excess of 95\%. This procedure allows for trivial linear transformations between the old and new CVs -- sign changes and rotations -- that do not affect the information content of the parameterization. If the boundary has not moved, the dimensionality has not changed, and the CVs have stabilized, then we terminate MESA. Otherwise, we perform a new round of interleaved CV discovery and biased sampling by cycling round the loop again from \blauw{Step 2}.

\textbf{Step 6 -- Compute free energy surface.} Perform umbrella sampling over the complete space spanned by the converged CVs specified by the terminal autoencoder. \blauwr{Construct the optimal estimate of the unbiased free energy surface in the collective variables $F(\boldsymbol\xi)$ by solving the WHAM equations, and optionally reweighting the unbiased free energy surface into arbitrary collective coordinates (\blauw{Section \ref{WHAM}}).}

\vspace{0.25in}

This algorithm possesses a number of attractive features. First, all steps are largely automated requiring minimal user intervention, including training of the autoencoder, boundary detection, and identification and launch of new umbrella runs. The requirement of the user beyond specifying the parameters of the algorithm is in \blauw{Step 5} to judge whether the boundary and/or CVs are sufficiently similar to warrant termination of MESA. In principle, this step could also be automated. Second, by interleaving CV discovery and enhanced sampling, we converge towards a set of global data-driven CVs providing a good parameterization of the accessible configurational space. Third, the availability of the explicit mapping from the atomic coordinates to the CVs enables enhanced sampling directly in the CVs. This stands in contrast to existing approaches that initialize unbiased simulations at the edge of the explored region in CV space to drive sampling of unexplored configurational space \cite{chiavazzo2017intrinsic,preto2014fast,zheng2013rapid}  and/or perform sampling in proxy physical variables\cite{ferguson2011integrating} or in an approximate basis function expansion \cite{hashemian2013modeling,li2006version,spiwok2011metadynamics,abrams2012fly,branduardi2007b}. Fourth, the collective coordinates are useful not only for enhanced sampling, but in providing understanding of the important slow collective motions governing the long-time evolution of the system. The explicit mapping between the atomic coordinates and CVs can help provide a transparent physical interpretation of these collective modes. Although these relations can be somewhat obscured by the complexity of the mapping function (\blauw{Eqn.~\ref{proj}}), the weights within the trained network can be quite informative in illuminating which atoms play important roles in each of the discovered CVs.

\subsection{\sffamily \large Molecular dynamics simulations}

\begin{figure}[ht!]
\begin{center}
\includegraphics[width=0.95\textwidth]{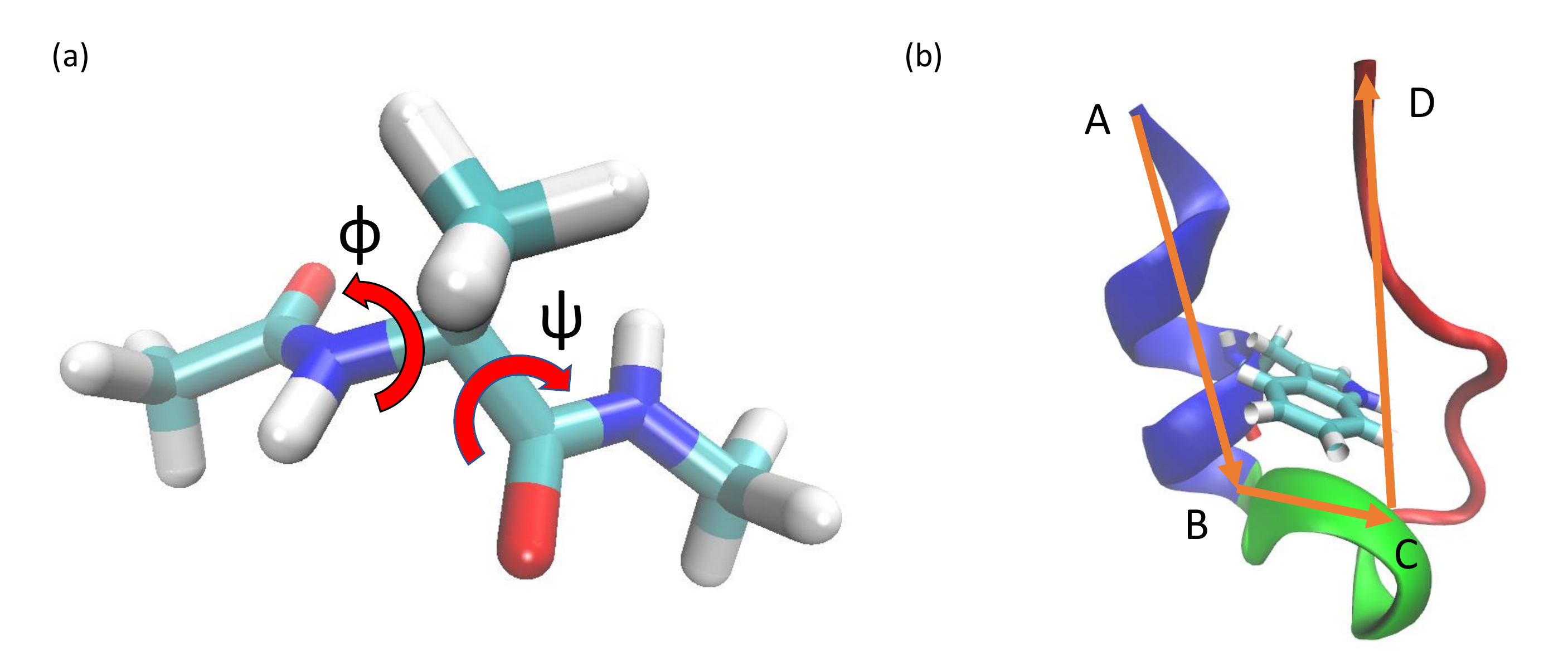}
\caption{Molecular structures of the two peptides studied in this work. (a) Alanine dipeptide indicating the $\phi$ and $\psi$ backbone dihedrals. (b) Trp-cage indicating the N-terminal $\alpha$-helix (blue), $3_{10}$ helix (green) and C-terminal polyproline region (red). The orange arrows denote the vectors linking the $\mathrm{C}_\alpha$ atoms of Asn-1 (A), Asp-9 (B), Ser-14 (C), and Ser-20 (D). The dihedral angle $\theta_{1,9,14,20}$ defined by these vectors measures the angle between the planes containing the $\mathrm{C}_\alpha$ atoms of \{Asn-1, Asp-9, Ser-14\}, and that containing those of \{Asp-9, Ser-14, Ser-20\}. This angle provides a measure of the global chirality of the molecule that is informative in understanding the CVs discovered by MESA. The Trp-6 side chain (explicitly visualized) is ``caged'' inside a hydrophobic core. All molecular renderings are constructed using VMD \cite{humphrey1996vmd}.}\label{AD}
\end{center}
\end{figure}

We demonstrate and validate MESA in applications to two peptides: alanine dipeptide and Trp-cage (\blauw{Fig.~\ref{AD}}). All molecular simulations were performed using the OpenMM 7.0 molecular dynamics package \cite{eastman2017openmm,eastman2012openmm,friedrichs2009accelerating}.

\subsubsection{\sffamily \normalsize Alanine dipeptide} \label{sec:methAla}

Simulations of alanine dipeptide (\textit{N}-acetyl-L-alanine-\textit{N}$^\prime$-methylamide, AcAlaNHMe) were performed in vacuum using the Amber99sb force field \cite{wang2004development}. Calculations were performed at $T$=300K and the equations of motion numerically integrated using a Langevin integrator with a friction coefficient of 1 ps$^{-1}$ and a 2 fs time step \cite{schlick2010molecular}. All bond lengths were fixed using the LINCS algorithm \cite{hess1997lincs}. Lennard-Jones interactions were treated with no cutoff, and we employed the Lorentz-Berthelot combining rules to determine interaction parameters between dissimilar atom types \cite{allenoxford}. Electrostatic interactions were computed exactly in real space using no cutoff. A single 800 ps initial unbiased calculation was performed, and snapshots saved every 1 ps to generate an initial unbiased ensemble of 800 molecular configurations. Enhanced sampling calculations were performed using an in-house biasing force plugin to OpenMM to launch $\nu$=10-20 umbrella windows in each round employing harmonic restraining potentials of $\kappa$=3000 kJ/mol.(unit of CV)$^2$ in each CV. Each umbrella calculation was performed for 100 ps, saving configurations every 1 ps. All calculations were performed on Intel i7-5820K chips (6-cores, 15 MB Cache, 3.8 GHz overclocked), achieving execution speeds of $\sim$6 $\mu$s/day.core for unbiased simulations and $\sim$1.3 $\mu$s/day.core for biased calculations.

\subsubsection{\sffamily \normalsize Trp-cage} \label{sec:methTrp}

Simulations of Trp-cage (NLYIQWLKDGGPSSGRPPPS; PDB ID: 1L2Y) were performed in a bath of 2773 TIP4P-Ew water molecules \cite{horn2004development} treating the protein using the Amber03 force field \cite{wang2004development}. A single Cl$^-$ counterion was added to neutralize the net single positive charge of Trp-cage. Initial molecular configurations were downloaded from the Protein Data-Bank \cite{neidigh2002designing,berman2002protein}. Calculations were conducted at $T$=300 K using an Andersen thermostat \cite{andersen1980molecular} and $P$=1 atm using a Monte-Carlo barostat\cite{chow1995isothermal,aaqvist2004molecular}. Equations of motion were numerically integrated using the leapfrog Verlet algorithm with a 2 fs time step \cite{hockney1988computer}. The length of all bonds involving H atoms were fixed using the LINCS algorithm \cite{hess1997lincs}. Lennard-Jones interactions were subjected to a 1 nm cutoff, and Lorentz-Berthelot combining rules used to determine interaction parameters between dissimilar atom types \cite{allenoxford}. Electrostatic interactions were treated using particle mesh Ewald with a 1 nm cutoff \cite{darden1993particle}. Three independent 10 ns unbiased simulations were conducted saving snapshots every 20 ps, to generate an initial unbiased ensemble of 1500 molecular configurations. Enhanced sampling calculations were performed using an in-house biasing force plugin to OpenMM to launch $\nu$=10-20 umbrella windows in each round employing harmonic restraining potentials of a minimum of $\kappa$=3000 kJ/mol.(unit of CV)$^2$ in each CV. Higher force constants were employed as necessary to achieve good sampling in rugged or steep regions of the landscape. Each umbrella calculation was performed for 2 ns, saving configurations every 20 ps. All calculations were performed on GeForce GTX 960 GPU cards, achieving execution speeds of $\sim$100 ns/day.core for the unbiased simulations and $\sim$95 ns/day.core for the biased calculations.

\subsection{\sffamily \large Implementation and open source availability}\label{implementation}

All molecular simulations were performed using the open source OpenMM 7.0 simulation package available for free download from \url{http://openmm.org} \cite{eastman2017openmm,friedrichs2009accelerating}. All artificial neural network construction and training was conducted using the Keras deep learning Python API available from \url{https://github.com/fchollet/keras} and running on top of the Theano libraries \cite{2016arXiv160502688full}.  Umbrella sampling calculations were performed using an in-house biasing force plugin \texttt{ANN\_Force} to OpenMM that computes all of the Jacobian matrix elements in \blauw{Eqn.~\ref{fiB}} from a trained autoencoder, and calculates from these the atomic biasing forces $\mathbf{f}_i^B$ for any given molecular configuration. We have implemented this plugin for both the CPU and CUDA platforms of OpenMM.

We have made all of our software implementations available for free and open source public download under the MIT License. The \texttt{ANN\_Force} biasing force plugin to OpenMM is available from \url{https://github.com/weiHelloWorld/ANN_Force}. The enhanced sampling framework to train autoencoders, perform boundary detection, deploy and run biased umbrella runs, and post-process the biased simulation data is available from \url{https://github.com/weiHelloWorld/accelerated_sampling_with_autoencoder}.

\section{\sffamily \Large Results and Discussion} \label{results_discussion_section}

We now discuss the application of MESA to two test peptides -- alanine dipeptide and Trp-cage -- to demonstrate, validate, and benchmark the approach (\blauw{Fig.~\ref{AD}}).

\subsection{\sffamily \large Alanine dipeptide}

Alanine dipeptide is a canonical and well-studied test system for new biomolecular methods \cite{hashemian2015topological,ferguson2011integrating,preto2014fast,zheng2013rapid,chodera2007use,hummer2003coarse,chodera2006long,chodera2007automatic} (\blauw{Fig.~\ref{AD}a}). It possesses a 2D intrinsic manifold that is conventionally parameterized by the $\phi$ and $\psi$ backbone dihedrals and lies on the surface of a flat torus \cite{hashemian2015topological,bolhuis2000reaction,preto2014fast,jakli2012variation}. The free energy surface supported by this intrinsic manifold is very well studied, and in vacuum supports a landscape containing three local minima corresponding to the $C_5$, $C_7$, and $\alpha_L$ states \cite{bolhuis2000reaction,preto2014fast}. Interconversion between the $C_5$ and $C_7$ states occurs on the order of $\sim$50 ps, but transitions to the $\alpha_L$ state are around 100-fold slower making efficient exploration and recovery of converged free energy surfaces a good test of our approach \cite{preto2014fast}. Unbiased and biased simulations of alanine dipeptide in vacuum were conducted as detailed in \blauw{Section \ref{sec:methAla}}.

\begin{figure}[ht!]
\begin{center}
\includegraphics[width=0.8\textwidth]{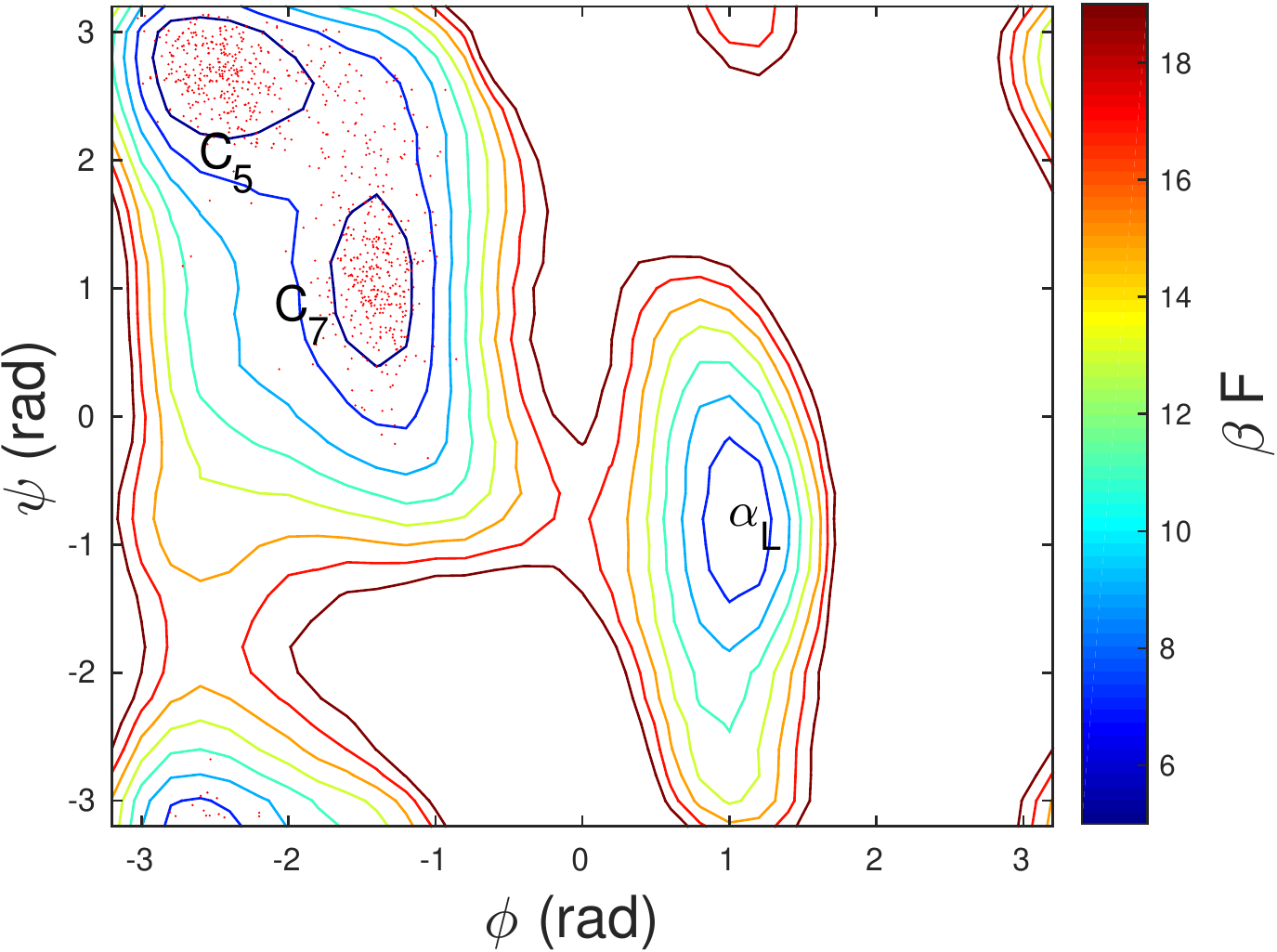} 
\caption{Projection into $\phi$-$\psi$ space of the initial 800 ps unbiased simulation trajectory used for the initial round of CV discovery. The underlying free energy landscape generated by direct umbrella sampling in $\phi$-$\psi$ space is plotted for reference. Contours of free energy $F$ are plotted in 2 $k_B T$ increments, where $\beta = k_B T$ is the reciprocal temperature, $k_B$ is Boltzmann's constant, and $T$=298 K. The unbiased trajectory (red points) initialized from the $C_5$ state rapidly explores the upper left corner containing the $C_5$ and $C_7$ basins, but fails to transition into the $\alpha_L$ state.}
\label{unbiased}
\end{center}
\end{figure}

\textbf{Generation of initial unbiased data (Step 1).} We commenced application of MESA by conducting a 800 ps unbiased simulation initialized from the $C_5$ state. As illustrated by the Ramachandran plot in \blauw{Fig.~\ref{unbiased}}, this long unbiased simulation comprehensively explores the $C_5$ and $C_7$ local minima, but fails to transition even once over the high free energy barriers separating them from the $\alpha_L$ state.

\textbf{Iterative CV discovery and sampling (Steps 2-5).} We then performed 10 rounds of interleaved CV discovery and enhanced sampling using MESA. Collective variable discovery was performed by training $D$-$G$-$K$-$G$-$2D$ autoencoders over $D$=21-dimensional input data $\textbf{z}$ comprising the Cartesian coordinates of the seven backbone atoms of alanine dipeptide. We employ two randomly selected reference configurations in our error function (\blauw{Eqn.~\ref{err_3}}), such that the output data are $2D$=42-dimensional. The number of nodes in the hidden layers was specified to be $G$=40 by cross validation. The number of bottleneck layer nodes in each iteration of MESA was determined by ascertaining the dimensionality of the intrinsic manifold by plotting the fraction of variance explained as a function of $K$ (\blauw{Eqn.~\ref{FVE}}). As illustrated in \blauw{Fig.~\ref{L_dim_Ala}} in the \blauw{Appendix}, we find a knee at $K$=2 in all 10 iterations, motivating the use of two bottleneck nodes in each loop of MESA. It is known that the intrinsic manifold of alanine dipeptide is 2D, so it is encouraging that our automated approach can accurately ascertain this from the data. Furthermore, it is known that the periodicity in the $\phi$ and $\psi$ dihedrals parameterizing the manifold endow it with the topology of a flat torus \cite{hashemian2015topological}. As carefully investigated by Hashemian and Arroyo \cite{hashemian2015topological}, topological periodicities in the intrinsic manifold can lead to inefficiencies in sampling and difficulties in CV determination. We shall see that these periodicities do not cause any such problems for MESA, but do introduce artifacts into the free energy surface. We quantify these effects below and present a protocol for their elimination.

We present in \blauw{Fig.~\ref{alanine_series}} the results of the 10 MESA iterations illustrating the expansion of the exploration of the 2D intrinsic manifold, stabilization of the CVs and explored region, and projection of sampling points into $\phi$-$\psi$ space. The explored region of the intrinsic manifold converges within 10 iterations, with the frontier of the intrinsic manifold and CVs stabilizing between iterations 9 and 10. The projections into $\phi$-$\psi$ space show that the system is driven to sample the $\alpha_L$ basin by iteration 6, indicating the ability of our approach to discover important collective motions and drive sampling in these collective coordinates.

\begin{figure}[ht!]
\begin{center}
\includegraphics[width=.85\textwidth]{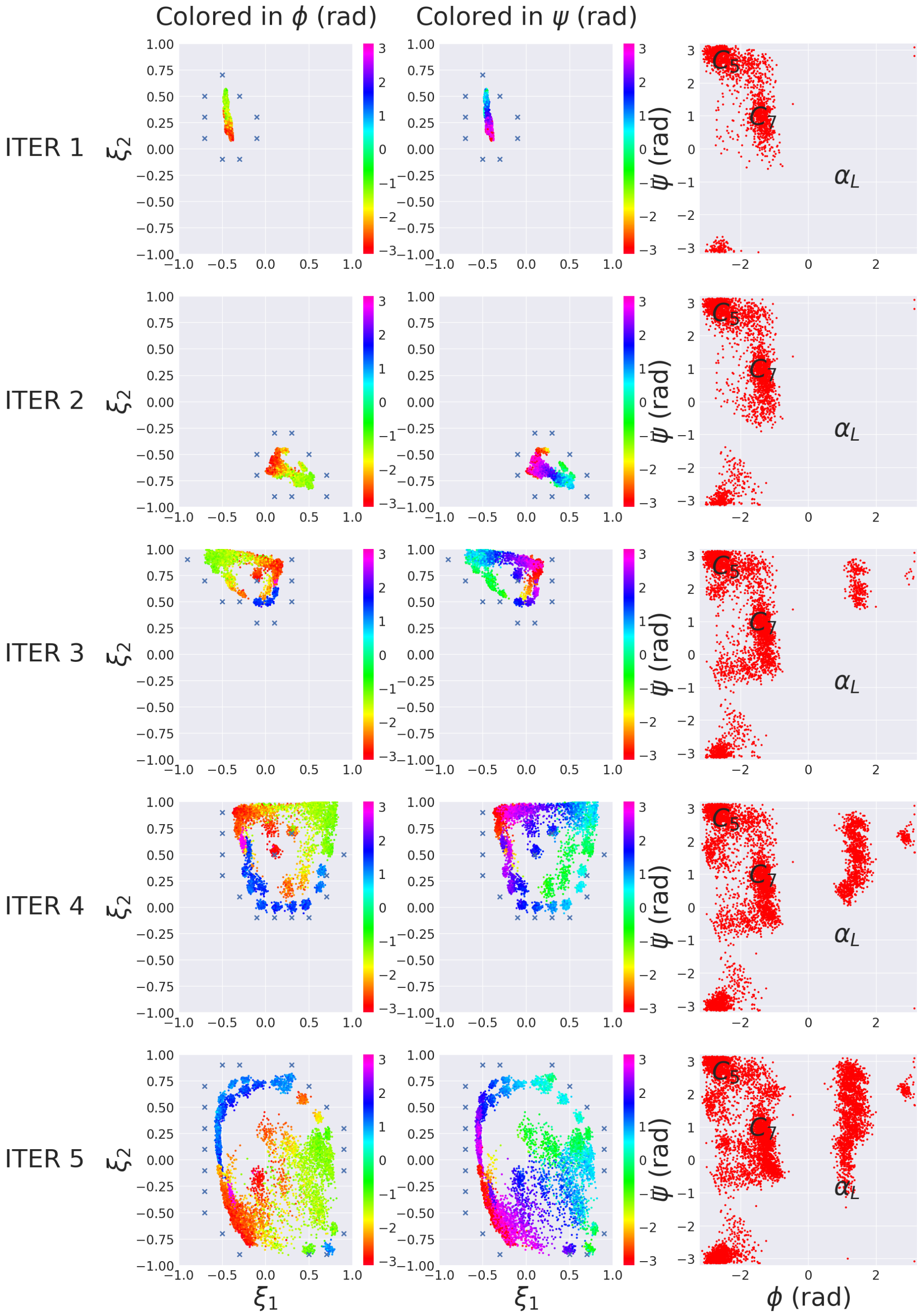} 
\end{center}
\end{figure}

\clearpage
\newpage

\begin{figure}[ht!]
\begin{center}
\includegraphics[width=.85\textwidth]{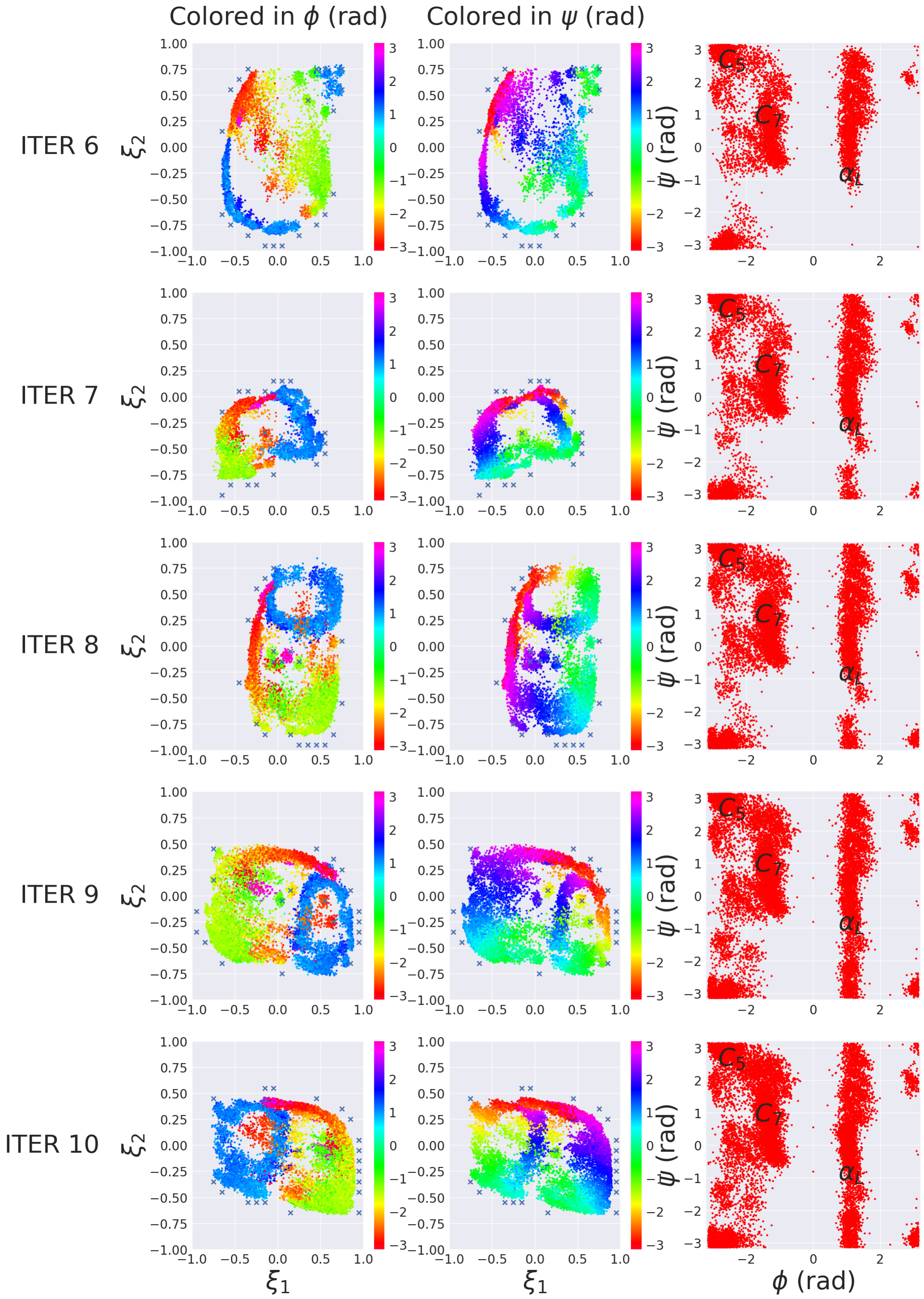} 
\end{center}
\end{figure}

\clearpage
\newpage

\begin{figure}[ht!]
\caption{Application of 10 rounds of MESA to alanine dipeptide in vacuum. Column 1 presents projections onto the intrinsic manifold of the current autoencoder spanned by $\{\xi_1,\xi_2\}$ all molecular configurations collected in the current iteration of biased sampling and all previous rounds of biased sampling. Points are colored by the $\phi$ dihedral angle. Crosses represent harmonic centers of umbrella sampling runs to be deployed in the current iteration that were identified by the boundary detection algorithm to advance the frontier of the explored region. Column 2 is identical to column 1, except points are colored by the $\psi$ dihedral angle. Column 3 presents Ramachandran plots containing the projection of molecular configurations into $\phi$-$\psi$ space. MESA rapidly drives sampling of the configurational space, and converges within 10 iterations. The biased sampling conducted in iteration 10 barely advances the frontier relative to iteration 9, and the CVs and explored regions are nearly identical up to a linear transformation.} \label{alanine_series}
\end{figure}

Inspection of the first two columns of \blauw{Fig.~\ref{alanine_series}} indicates that there is no simple monotonic relationship between the autoencoder discovered CVs $\{\xi_1,\xi_2\}$ and the backbone dihedrals $\{\phi,\psi\}$ known to provide a good parameterization of the intrinsic manifold. The source of this apparent discrepancy is the inherent periodicity in these two dihedral angles, which prevents the emergence of a simple bijective mapping with the autoencoder CVs. Close inspection of \blauw{Fig.~\ref{alanine_series}} reveals the emergence of closed loops in $\psi$ space corresponding to wrapping of the enhanced sampling through the $\psi = \pi / -\pi$ periodic boundary. Accordingly, the intrinsic manifold spanned by $\xi_1$-$\xi_2$ actually represents a projection of the topologically closed flat torus upon which the alanine dipeptide intrinsic manifold resides \cite{hashemian2015topological}. Indeed, the flat torus can only be embedded in four dimensions or higher \cite{patrangenaru2015nonparametric}. Mathematically, four dimensions are required to preserve the topological structure of the flat torus and define an invertible map such that every point on the surface of the torus is uniquely located within the projection and the object is fully unfolded in the embedding without artificial intersections \cite{patrangenaru2015nonparametric,whitney1936differentiable,broomhead1986extracting}. We should therefore expect projections into lower dimensional spaces to contain regions where distant points on the flat torus are projected to identical points in the low-dimensional embedding \cite{hashemian2015topological}. Such artifacts are visually apparent in the last row of \blauw{Fig.~\ref{alanine_series}}, where the upper right quadrant of the blue $\phi$ $\approx$ 1 rad loop containing configurations with [$\phi$ $\approx$ 1 rad, $\psi$ $\approx$ 2 rad] appears to lie on top of configurations with [$\phi$ $\approx$ -2.5 rad, $\psi$ $\approx$ 0 rad]. In the present application these artifacts do not appear to be so severe as to introduce any difficulties for CV determination or enhanced sampling using MESA, but we do observe that they can impede sampling efficiency by causing exploration to become stuck within these loops for a few iterations before MESA is able to drive sampling in the other degree of freedom.

\begin{figure}[ht!]
\begin{center}
\includegraphics[width=0.95\textwidth]{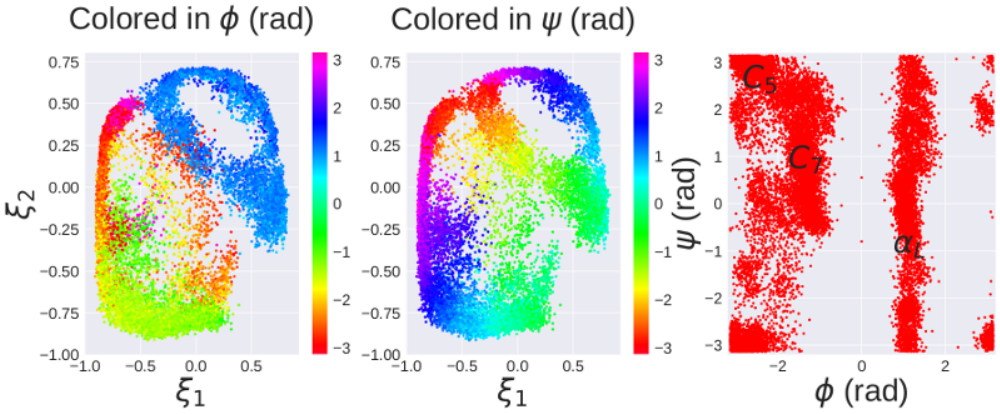} 
\caption{Projections of all molecular snapshots harvested over the course of the 10 MESA iterations into the 2D intrinsic manifold spanned by the CVs $\{\xi_1,\xi_2\}$ identified by the terminal autoencoder. (left) Projection into the 2D intrinsic manifold colored by the $\phi$ dihedral angle. (center) Projection into the 2D intrinsic manifold colored by the $\psi$ dihedral angle. (right) Ramachandran plot projection into $\phi$-$\psi$ space.}
\label{2D_autoencoder_plot_used_in_WHAM}
\end{center}
\end{figure}

\textbf{Free energy surface (Step 6).} Having converged MESA, we collate all simulation snapshots collected over the 10 iterations to train a terminal autoencoder that will be used to perform umbrella sampling over the complete CV space. An analogous plot to \blauw{Fig.~\ref{alanine_series}} for this terminal autoencoder is presented in \blauw{Fig.~\ref{2D_autoencoder_plot_used_in_WHAM}}. 

Using this terminal autoencoder, we conduct 200 umbrella sampling calculations applying biasing potentials to tile the intrinsic manifold spanned by the CVs $\{\xi_1,\xi_2\}$. We then analyze these biased trajectories to estimate the unbiased free energy surface by solving the WHAM equations (\blauw{Section \ref{WHAM}}). We present in \blauw{Fig.~\ref{2D_FES}a} the free energy surface $F(\xi_1,\xi_2)$ in CV space and in \blauw{Fig.~\ref{2D_FES}b} the estimated uncertainties in the landscape computed by 10 rounds of bootstrap resampling with replacement. In \blauw{Fig.~\ref{2D_FES}c-d} we present analogous plots for $F(\phi,\psi)$ and its estimated uncertainties computed by reweighting of our biased simulation data collected in $\{\xi_1,\xi_2\}$. As a point of comparison, we present in \blauw{Fig.~\ref{2D_FES}e-f} an estimate of $F(\phi,\psi)$ and its uncertainties computed by performing umbrella sampling directly in the dihedral angles $\{\phi,\psi\}$ as a ground truth estimate of the free energy surface. Finally, we present in \blauw{Fig.~\ref{2D_FES}g} a plot of the difference between the optimally aligned $F(\phi,\psi)$ landscapes computed by reweighting umbrella sampling data in $\{\xi_1,\xi_2\}$ (\blauw{Fig.~\ref{2D_FES}c}) and direct umbrella sampling in $\{\phi,\psi\}$ (\blauw{Fig.~\ref{2D_FES}e}). Analysis of \blauw{Fig.~\ref{2D_FES}} reveals that MESA ably resolves the basins corresponding to the three metastable states $C_5$, $C_7$, and $\alpha_L$, but \blauw{Fig.~\ref{2D_FES}g} shows that the shape, location and depth of the $\alpha_L$ basin differs from that recovered from direct umbrella sampling in $\{\phi,\psi\}$, and a spurious metastable minimum is resolved at [$\phi$ $\approx$ -2.5 rad, $\psi$ $\approx$ -0.75 rad].  These artifacts in the free energy surface are a direct result of the periodicities in the underlying intrinsic manifold that populates the surface of a flat torus and cannot be correctly embedded in a 2D projection \cite{hashemian2015topological}.

\begin{figure}[ht!]
\begin{center}
\includegraphics[width=.99\textwidth]{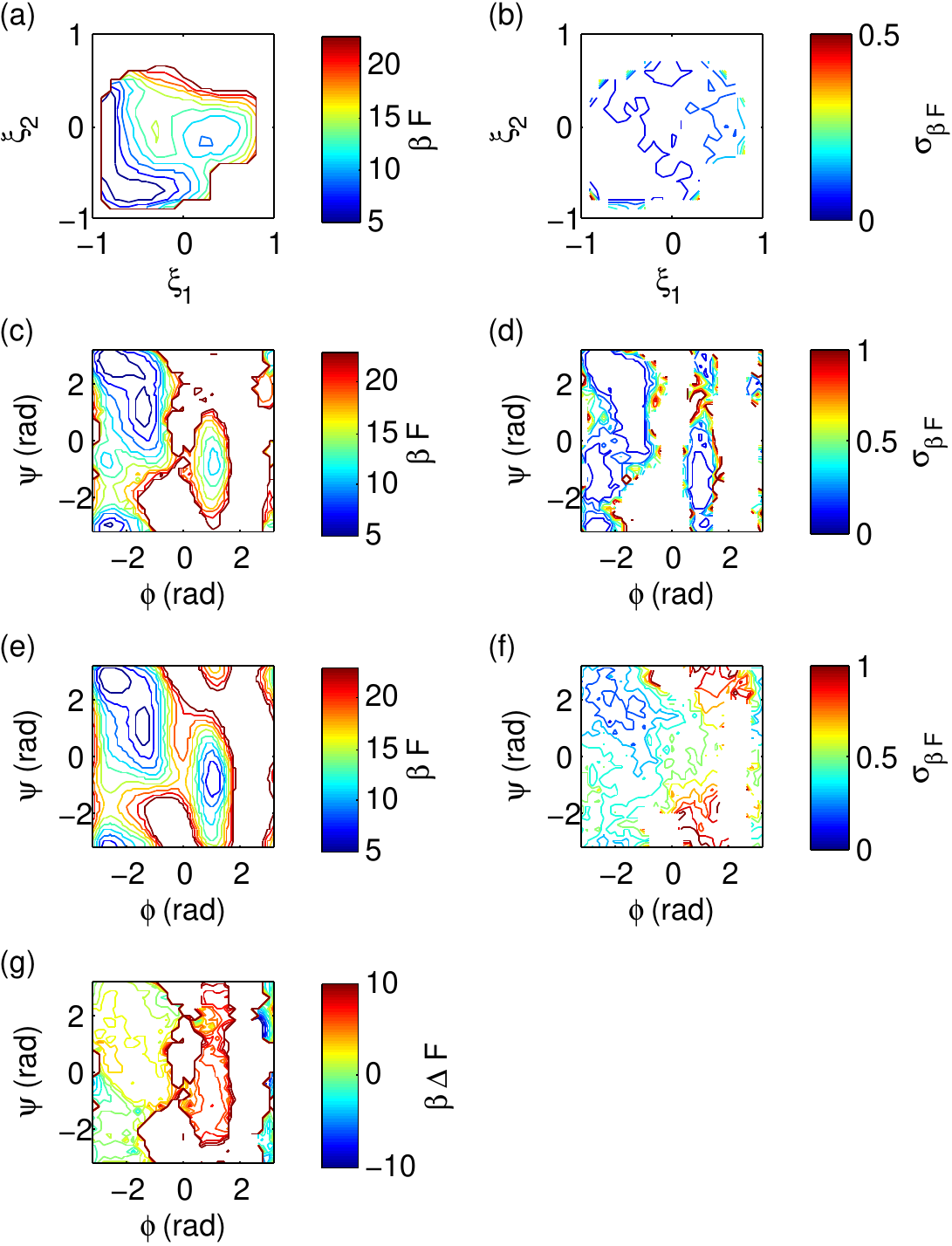} 
\end{center}
\end{figure}

\clearpage
\newpage

\begin{figure}[ht!]
\caption{Comparison of free energy surfaces of alanine dipeptide in vacuum obtained from 2D biased sampling in MESA CVs and direct sampling in the $\{\phi,\psi\}$ dihedrals. (a,b) Free energy surface $F(\xi_1,\xi_2)$ and associated estimated uncertainties $\sigma_{\beta F}$ computed by biased sampling in MESA identified CVs. Free energies are reported in units of $k_B T$, where $\beta = 1/k_B T$, $k_B$ is Boltzmann's constant and $T$=298 K. (c,d) Free energy surface $F(\phi,\psi)$ and associated uncertainties $\sigma_{\beta F}$ computed by reweighting the biased simulation data collected in $\{\xi_1,\xi_2\}$. (e,f) Free energy surface $F(\phi,\psi)$ and associated uncertainties $\sigma_{\beta F}$ computed by conducting umbrella sampling directly in the dihedral angles $\{\phi,\psi\}$. Uncertainties in the landscapes are computed by 10 rounds of bootstrap resampling with replacement. (g) Free energy difference landscape $\Delta F\{\phi,\psi\}$ formed by optimally aligning and then subtracting the landscape in (e) from that in (c). Biased sampling in the MESA CVs misidentifies the location, shape, and depth of the $\alpha_L$ well, and identifies a spurious metastable minimum at [$\phi$ $\approx$ -2.5 rad, $\psi$ $\approx$ -0.75 rad].}
\label{2D_FES}
\end{figure}

A straightforward way to solve the projection problem with periodic variables is to simply increase the dimensionality of the autoencoder nonlinear projection by increasing the number of bottleneck layer nodes. The Whitney Embedding Theorem states that any $n$ dimensional manifold can be embedded (i.e., fully unfolded without spurious crossings) in $\mathbb{R}^d$ for $d$ = (2$n$+1) \cite{whitney1936differentiable}. The theorem specifies the embedding dimensionality at or above which we are guaranteed to achieve a proper embedding, but depending on the topology of the manifold it may be possible to employ $d$ $<$  (2$n$+1). Accordingly, if we have prior knowledge that the intrinsic manifold of our system may contain periodicities, or we observe the emergence of closed loops in the CVs identified by MESA (e.g., \blauw{Fig.~\ref{alanine_series}}), then we propose that the number of hidden nodes be increased to (2$K$+1), where $K$ is the dimensionality of the intrinsic manifold identified by the FVE approach described in \blauw{Section \ref{AE}}. Increasing dimensionality does come with increasing computational cost since it is more expensive to train the autoencoder (since the number of network parameters increases), run biased calculations (since the analytical expression for the nonlinear projection becomes more complicated), and conduct umbrella sampling (due to the curse of dimensionality). It is also more challenging to interpret and visualize data high-dimensional spaces. In practice, therefore, we suggest that the number of hidden nodes may be increased one at a time and the dimensionality increase terminated when projections of the manifold cease to depend on $K$.

\begin{figure}[ht!]
\begin{center}
\includegraphics[width=\textwidth]{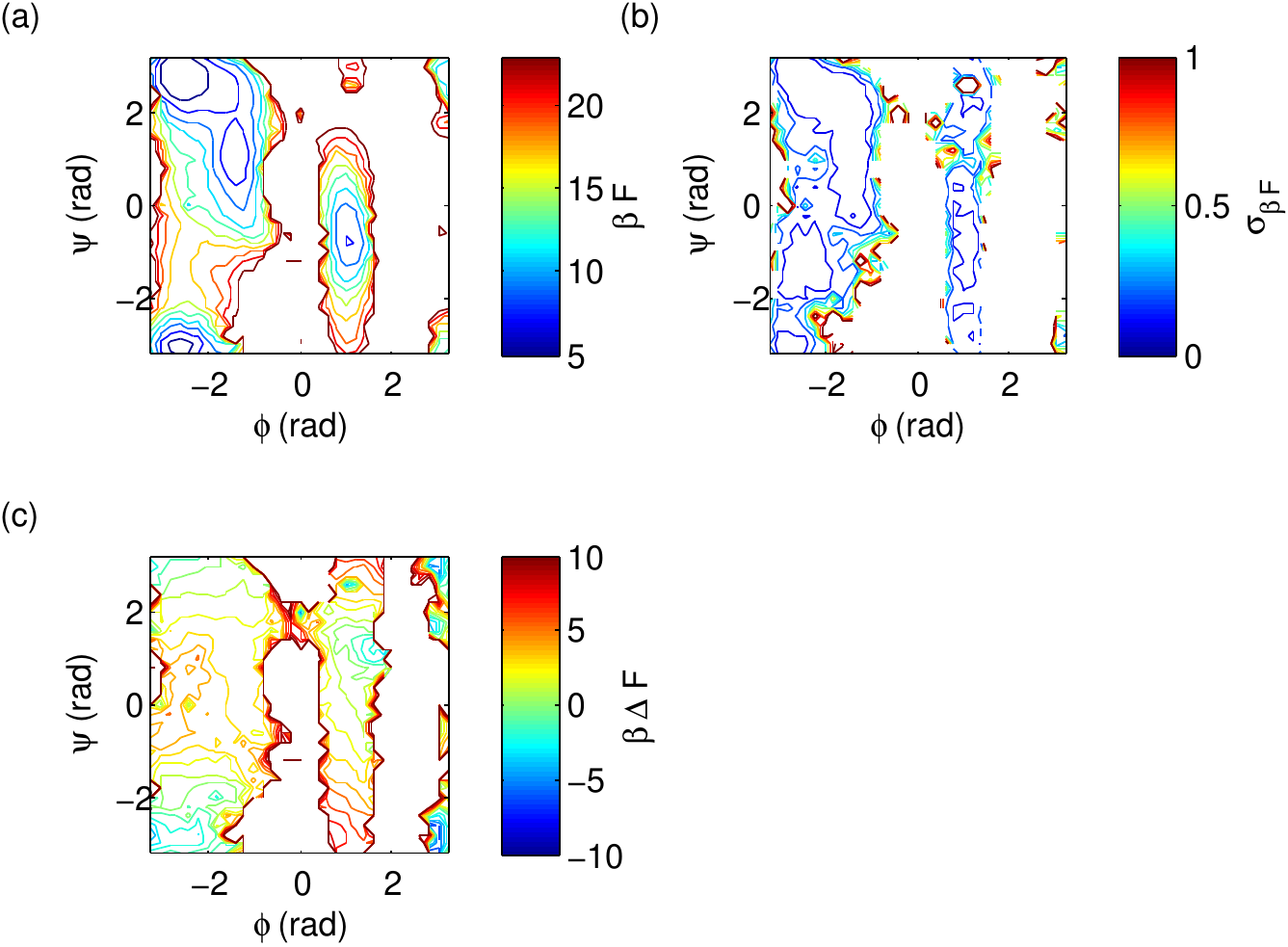} 
\caption{Comparison of free energy surfaces of alanine dipeptide in vacuum obtained from 4D biased sampling in MESA CVs and direct sampling in $\{\phi,\psi\}$ space.  (a,b) Free energy surface $F(\phi,\psi)$ and associated uncertainties $\sigma_{\beta F}$ computed by reweighting the biased simulation data collected in $\{\xi_1,\xi_2, \xi_3,\xi_4\}$. Free energies are reported in units of $k_B T$, where $\beta = 1/k_B T$, $k_B$ is Boltzmann's constant and $T$=298 K. (c) Free energy difference landscape $\Delta F\{\phi,\psi\}$ formed by optimally aligning and then subtracting the landscape in \blauw{Fig.~\ref{2D_FES}e} from that in (a). Sampling in MESA identified CVs in 4D fully unfolds the intrinsic manifold of alanine dipeptide, and the projected FES into $\phi$-$\psi$ shows excellent agreement with that recovered by direct sampling in these dihedral angles.}
\label{4D_FES}
\end{center}
\end{figure}

In the present case, we know that $\mathbb{R}^4$ is sufficient to embed the flat torus containing the intrinsic manifold of alanine dipeptide \cite{hashemian2015topological}, and so we train a new terminal autoencoder with $K$=4 bottleneck nodes over the collated data from all 10 rounds of MESA conducted using $K$=2 topologies. We then performed 1000 umbrella sampling simulations to tile the 4D space spanned by the autoencoder CVs $\{\xi_1,\xi_2,\xi_3,\xi_4\}$ and recover an estimate of the unbiased FES by solving the WHAM equations. In \blauw{Fig.~\ref{4D_FES}a} we present $F(\phi,\psi)$ computed by reweighting the the umbrella sampling calculations, and in \blauw{Fig.~\ref{4D_FES}b} the estimated uncertainties computed by 10 rounds of bootstrap resampling with replacement. In \blauw{Fig.~\ref{4D_FES}c}, we present the difference between \blauw{Fig.~\ref{4D_FES}a} and that recovered by direct umbrella sampling in $\{\phi,\psi\}$ (\blauw{Fig.~\ref{2D_FES}e}). By fully expanding the flat torus into 4D space, we eliminate the projection problem associated with periodic variables. MESA ably identifies four CVs parameterizing the embedding of the flat torus, and projection of the 4D free energy surface into $\phi$-$\psi$ space shows excellent agreement with that recovered by direct sampling in these dihedral angles.

\subsection{\sffamily \large Trp-cage}

Trp-cage (PDB ID: 1L2Y) is a 20-residue mini-protein \cite{neidigh2002designing} with a $\sim$4 $\mu$s folding time and a native state containing an $\alpha$-helix, a $3_{10}$-helix, polyproline II helix, and salt bridge that ``cage'' Trp-6 in a hydrophobic core \cite{kim2015systematic,qiu2002smaller,seshasayee2005high,heyda2011urea,juraszek2006sampling} (\blauw{Fig.~\ref{AD}b}). A fast-folding mini-protein containing secondary structural elements present in large proteins, Trp-cage has been extensively studied by both computation\cite{kim2015systematic,kim2016computational,juraszek2006sampling,hatch2014computational,kannan2014role,schug2005energy,snow2002trp,zhou2003trp,zhou2003free,deng2013kinetics,patriksson2007direct,pitera2003understanding} and experiment\cite{qiu2002smaller,meuzelaar2013folding,barua2008trp,iavarone2005conformational,rovo2013structural,adams2006probing} as the ``hydrogen atom of protein folding'' \cite{heyda2011urea,bolhuis2009two}. In particular, Jurazek and Bolhuis identified two distinct folding pathways using transition path sampling \cite{juraszek2006sampling}. Deng \textit{et al.}\ used Markov state models to analyze an long 208 $\mu$s simulation to resolve similar patterns in overall structure changes and folding routes \cite{deng2013kinetics}. Kim \textit{et al.}\ ran dozens of folding simulations starting from unfolded states, then projected these trajectories into a 2D collective variable space identified by diffusion maps\cite{kim2015systematic}.  Both Deng \textit{et al.}\ and Kim \textit{et al.}\ achieved good agreement of their estimated folding times with experiment, demonstrating the ability of MD simulations to quantitatively recapitulate biophysical folding processes. We employ Trp-cage as a more realistic and challenging test case for MESA. Simulations of Trp-cage in explicit water were conducted as detailed in \blauw{Section \ref{sec:methTrp}}.

\textbf{Generation of initial unbiased data (Step 1).}  We ran three independent 10 ns unbiased simulations commencing from the Trp-cage native state downloaded from the Protein Data-Bank (PDB ID: 1L2Y) \cite{neidigh2002designing,berman2002protein}. Snapshots were saved every 20 ps to generate an initial unbiased training set of 1500 molecular configurations. These data achieve a maximum root mean squared deviation (\groen{RMSD}) of the $\mathrm{C}_\alpha$ atoms relative to the native state of 3.5 \AA, and the first iteration of MESA results in an autoencoder with an FVE $\approx$ 60\%. Accordingly, these unbiased data provide a reasonable starting ensemble for iterative application of MESA.

\textbf{Iterative CV discovery and sampling (Steps 2-5).} We then performed six rounds of interleaved CV discovery and enhanced sampling using MESA before achieving convergence. Collective variable discovery was performed by training $D$-$G$-$K$-$G$-$2D$ autoencoders over $D$=180-dimensional input data $\textbf{z}$ comprising the Cartesian coordinates of the 60 backbone atoms of Trp-cage. We again used two randomly selected reference configurations in our error function (\blauw{Eqn.~\ref{err_3}}), so that the output data are 2$D$=360-dimensional. The number of nodes in the hidden layers were tuned to $G$=50 by cross-validation. The number of bottleneck layer nodes in each iteration of MESA by plotting the FVE as a function of $K$ using \blauw{Eqn.~\ref{FVE}}. As illustrated in \blauw{Fig. \ref{L_dim_Trp_cage}} in the \blauw{Appendix}, we find a knee at $K$=2 in all 6 iterations, supporting the use of two bottleneck nodes in each iteration of MESA. This estimated dimensionality of the Trp-cage intrinsic manifold is consistent with prior work \cite{juraszek2006sampling,kim2015systematic,zhou2003trp}. 

We present in \blauw{Fig.~\ref{Trp_cage_series}} the results of the six MESA iterations illustrating accelerated sampling and exploration of the intrinsic manifold. We project the sampled molecular configurations into the 2D intrinsic manifold spanned by the two MESA CVs $\{\xi_1,\xi_2\}$ discovered in each iteration and colored according to three physical order parameters to illuminate the configurational diversity in the accelerated sampling: the RMSD of the $\mathrm{C}_\alpha$ atoms relative to the native state, the end-to-end extent of the molecule $d_{1,20}$ measured between the $\mathrm{C}_\alpha$ atoms in Asn-1 and Ser-20, and the sine of the non-local dihedral angle $\theta_{1,9,14,20}$ formed by the $\mathrm{C}_\alpha$ atoms in Asn-1, Asp-9, Ser-14, and Ser-20. As illustrated in \blauw{Fig.~\ref{AD}b}, $\theta_{1,9,14,20}$ measures the signed angle between the planes containing the $\mathrm{C}_\alpha$ atoms of \{Asn-1, Asp-9, Ser-14\}, and that containing those of \{Asp-9, Ser-14, Ser-20\}. $\mathrm{C}_\alpha$-RMSD measures the departure of the configurational sampling from the native state, $d_{1,20}$ measures the linear extent of the peptide chain, and $\sin \theta_{1,9,14,20}$ measures the global chirality of the molecule: $\sin \theta_{1,9,14,20}>0$ indicates that the N-terminal and C-terminal strands are arranged such that the molecule adopts a right-handed chirality, $\sin \theta_{1,9,14,20}<0$ that they adopt a left-handed chirality, and $\sin \theta_{1,9,14,20}$=0 that the N-terminal and C-terminal strands lie in the same plane. Convergence of the frontier of the intrinsic manifold, volume of configurational space explored, and stabilization of the identified CVs is achieved between iterations 5 and 6.

\begin{figure}[ht!]
\begin{center}
\includegraphics[width=\textwidth]{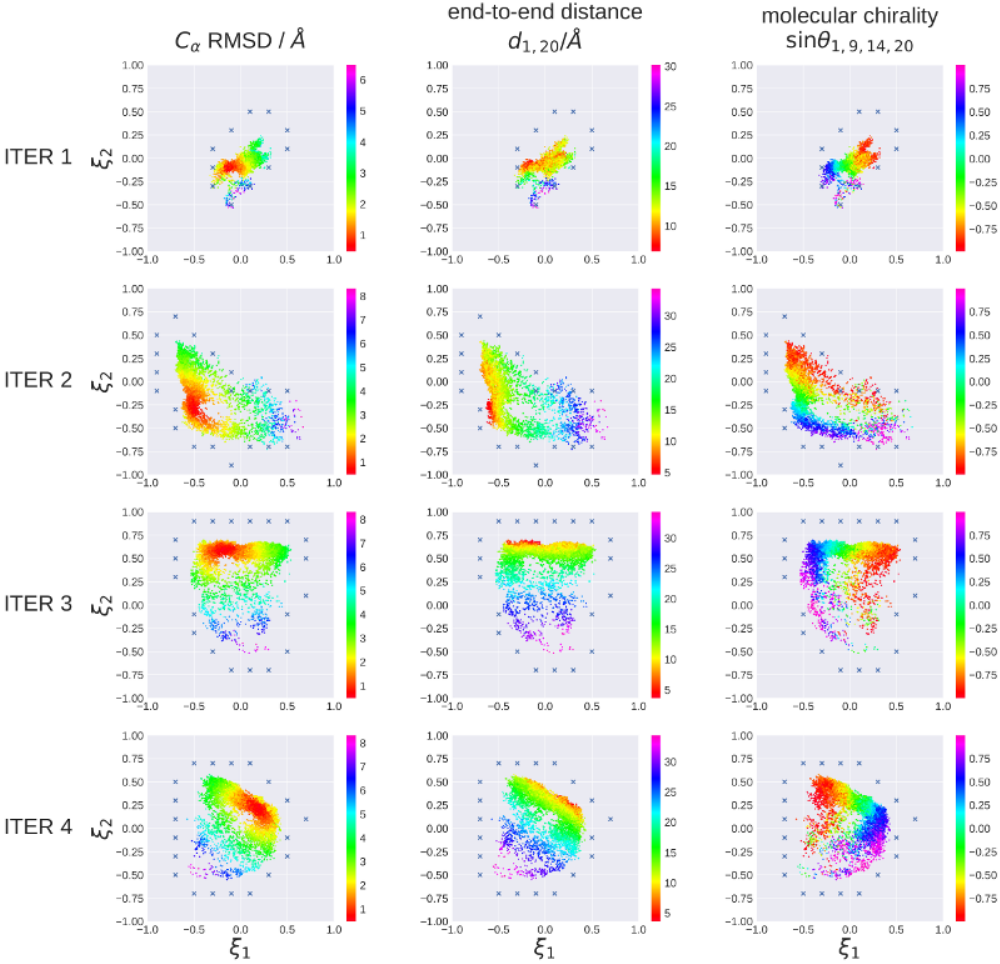} 
\end{center}
\end{figure}

\clearpage
\newpage

\begin{figure}[ht!]
\begin{center}
\includegraphics[width=\textwidth]{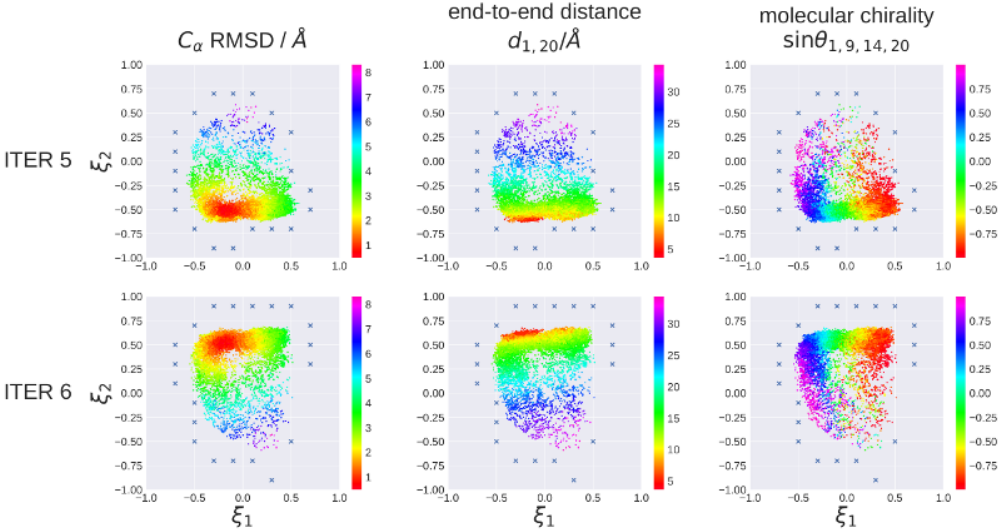} 
\caption{Application of six rounds of MESA to Trp-cage in explicit solvent. Each row corresponds to a particular MESA iteration in which all molecular configurations collected in the current and previous rounds of biased sampling are projected as points into the 2D intrinsic manifold spanned by the current CVs $\{\xi_1,\xi_2\}$. Each column presents a different coloration of the projected data by selected physical order parameters that illuminate the increasing configurational diversity in each successive MESA iteration: the $\mathrm{C}_\alpha$ RMSD (column 1), the end-to-end distance $d_{1,20}$ measured between the $\mathrm{C}_\alpha$ atoms in Asn-1 and Ser-20 (column 2), and the non-local dihedral angle $\theta_{1,9,14,20}$ formed by the $\mathrm{C}_\alpha$ atoms in Asn-1, Asp-9, Ser-14, and Ser-20. Crosses represent harmonic centers of umbrella sampling runs to be deployed in the current iteration that were identified by the boundary detection algorithm. MESA rapidly drives sampling of the accessible configurational space, and converges within 6 iterations. Umbrella sampling barely advances the frontier between iterations 5 and 6, and the CVs and explored regions are nearly identical up to a linear transformation.} \label{Trp_cage_series}
\end{center}
\end{figure}

\begin{figure}[ht!]
\begin{center}
\includegraphics[width=\textwidth]{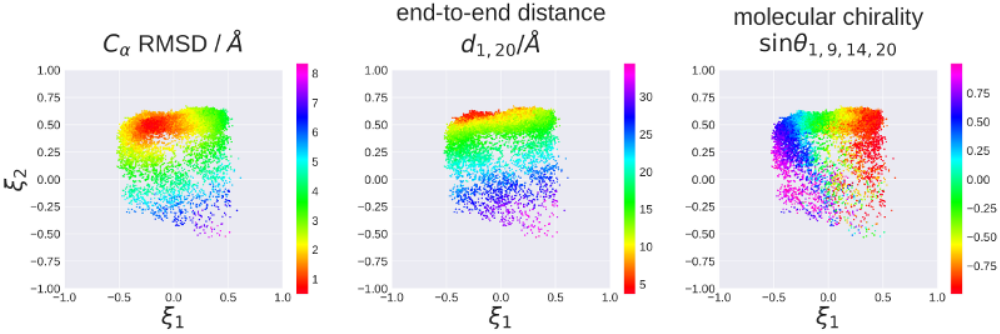} 
\caption{Projections of all molecular snapshots harvested over the course of the six MESA iterations iterations into the 2D intrinsic manifold spanned by the CVs $\{\xi_1,\xi_2\}$ identified by the terminal autoencoder. Points are colored by (left) $\mathrm{C}_\alpha$-RMSD, (center) $d_{1,20}$, and (right) $\theta_{1,9,14,20}$. }
\label{Trp_WHAM}
\end{center}
\end{figure}

\textbf{Free energy surface (Step 6).} Upon convergence of MESA, we compile all of the simulation snapshots collected over the six iterative cycles to train a terminal autoencoder with which to perform umbrella sampling over the intrinsic manifold. An analogous plot to \blauw{Fig.~\ref{Trp_cage_series}} for this terminal autoencoder is presented in \blauw{Fig.~\ref{Trp_WHAM}}. 

\begin{figure}[ht!]
\begin{center}
\includegraphics[width=0.90\textwidth]{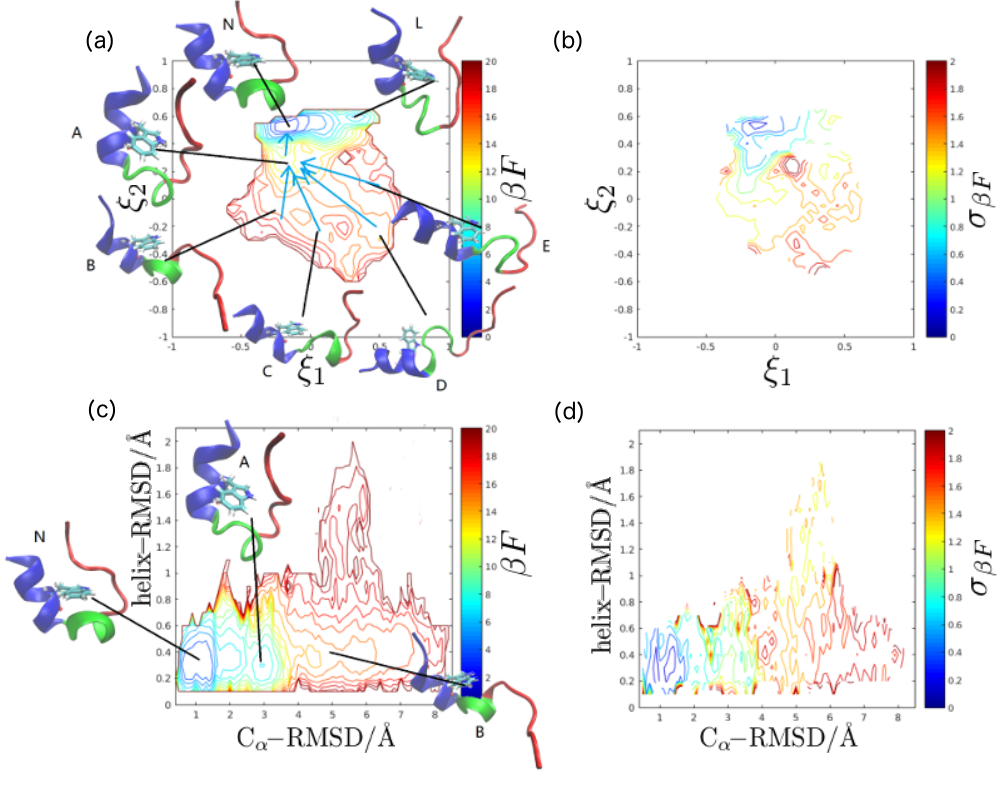} 
\caption{Free energy surfaces for Trp-cage in explicit solvent obtained from 2D biased sampling in MESA CVs. (a,b) Free energy surface $F(\xi_1,\xi_2)$ and associated estimated uncertainties $\sigma_{\beta F}$ computed by biased sampling in MESA identified CVs $\{\xi_1,\xi_2\}$. (c,d) Free energy surface $F$($\mathrm{C}_\alpha$-RMSD, helix-RMSD) and associated uncertainties $\sigma_{\beta F}$ computed by reweighting the biased simulation data collected in $\{\xi_1,\xi_2\}$. $\mathrm{C}_\alpha$-RMSD is the RMSD of all $\mathrm{C}_\alpha$ atoms in the peptide from the Trp-cage native fold. helix-RMSD is the RMSD of the $\mathrm{C}_\alpha$ atoms in the N-terminal $\alpha$-helix (residues 2-8) from the Trp-cage native fold. Free energies are reported in units of $k_B T$, where $\beta = 1/k_B T$, $k_B$ is Boltzmann's constant and $T$=298 K. Uncertainties in the landscapes are computed by 10 rounds of bootstrap resampling with replacement. Superposed on the landscape are representative molecular configurations in which the N-terminal $\alpha$-helix region is colored in blue, the $3_{10}$ helix in green, and the C-terminal polyproline region in red, and the Trp-6 side chain explicitly rendered.}
\label{FES_Trp_cage}
\end{center}
\end{figure}

We then perform 30 umbrella sampling runs tiling the explored region of the intrinsic manifold spanned by the converged CVs $\{\xi_1,\xi_2\}$ identified by the terminal autoencoder. Finally, we estimate the unbiased free energy surface by solving the WHAM equations (\blauw{Section \ref{WHAM}}). We present in \blauw{Fig.~\ref{FES_Trp_cage}a} the free energy surface $F(\xi_1,\xi_2)$ and in \blauw{Fig.~\ref{FES_Trp_cage}b} the estimated uncertainties in the landscape computed by 10 rounds of bootstrap resampling with replacement. We superpose onto the landscape representative molecular configurations to illustrate the folding pathways, and also mark the native state (\textbf{N}) and a partially unfolded state (\textbf{L}) that is similar to the previously reported ``loop structure'' in which the $3_{10}$ helix is unfolded while the other secondary structure elements remain intact\cite{kim2015systematic,juraszek2006sampling,kannan2009folding,shao2012enhanced}.  Comparison of \blauw{Fig.~\ref{FES_Trp_cage}a} with \blauw{Fig.~\ref{Trp_WHAM}} reveals a physical interpretation of the CVs $\{\xi_1,\xi_2\}$. We find $\xi_1$ to be strongly correlated with the global molecular chirality $\sin \theta_{1,9,14,20}$ (Pearson correlation coefficient $\rho$ = -0.92 with 2-tailed p-value $<$ 10$^{-15}$), indicating that MESA has discovered a CV that distinguishes left-handed, right-handed, and planar molecular configurations. Meanwhile, $\xi_2$ is strongly correlated with both the end-to-end distance $d_{1,20}$ (Pearson correlation coefficient $\rho$ = -0.91 with 2-tailed p-value $<$ 10$^{-15}$) indicating that MESA identifies a second CV tracking the global linear extent of the peptide. 

These two CVs describing the important collective motions of Trp-cage map out a rugged 2D folding landscape and resolve a low-free energy folding pathway indicated by the blue arrows in \blauw{Fig.~\ref{FES_Trp_cage}a}. Starting from extended configurations with an intact N-terminal helix, fully unfolded $3_{10}$ helix, and different global chiralities -- states \textbf{B, C, D, E} -- these states move along shallow valleys within a relatively flat free energy plateau to reach an intermediate state \textbf{A} where the hydrophobic core and a $\sim$7 \AA\ salt bridge between Asp-9 and Arg-16 is formed. In this configuration, Trp-6 is buried in the hydrophobic core and the free energy of the system by 3-7 $k_B T$ lower than the plateau. Intermediate state \textbf{A} is structurally quite similar to native state, possessing a $C_\alpha$ RMSD of $\sim$3 \AA\ and global molecular planarity. The final transition to the native state \textbf{N} proceeds down a steep free energy valley corresponding to folding of $3_{10}$ helix and a further $\sim 5k_B T$ decrease in free energy.

The loop structure \textbf{L} represents a low-lying metastable state that is only $\sim$2 $k_B T$ less stable than the native fold, and accessible over a $\sim$2 $k_B T$ free energy barrier. Transformation of the native fold into the loop structure is accompanied by unfolding of the $3_{10}$ helix and the concomitant adoption of a left-handed chirality. In results reported by Juraszek \textit{et al.} \cite{juraszek2006sampling} and Kim \textit{et al.}\cite{kim2015systematic}, \textbf{L} state is an important intermediate in the nucleation-condensation folding path. 

In \blauw{Fig.~\ref{FES_Trp_cage}c} we project our FES into the two physical order parameters ($\mathrm{C}_\alpha$-RMSD, helix-RMSD) -- where helix-RMSD is the RMSD of the $\mathrm{C}_\alpha$ atoms in the N-terminal $\alpha$-helix (residues 2-8) from the Trp-cage native fold -- in order to draw comparisons with landscapes and folding pathways previously reported by Juraszek \textit{et al.}\ (Fig.\ 3) \cite{juraszek2006sampling}, Kim \textit{et al.}\ (Fig.\ 2) \cite{kim2015systematic}, Paschek \textit{et al.}\ (Fig.\ 3) \cite{paschek2008computing}, and Deng \textit{et al.}\ (Fig.\ 5) \cite{deng2013kinetics}. The FES mapped out in these order parameters by MESA shows good agreement with that computed by replica exchange molecular dynamics (\groen{REMD}) by Juraszek \textit{et al.}\ employing the OPLSAA force field \cite{kaminski2001evaluation} and SPC water model\cite{lindahl2001gromacs}. Our folding pathway also closely resembles the diffusion-collision \textbf{I}-\textbf{P}$_d$-\textbf{N} folding path resolved in that study using transition path sampling (\groen{TPS}). Conversely, in our calculations the N-terminal $\alpha$-helix is always intact, whereas Juraszek \textit{et al.}\ observe it to unfold, opening up a second nucleation-condensation folding pathway via the \textbf{L} state. Interestingly, this folding path is only observable using TPS, and proceeds through regions of the FES unmapped by the REMD calculations, which -- as the authors observe -- had difficulty surmounting high free energy barriers and may not be fully converged. We note that preservation of $\alpha$-helical content in the unfolded state is consistent with spectroscopy measurements conducted by Ahmed \textit{et al.}\ \cite{ahmed2005uv} suggesting early formation of the $\alpha$-helix \cite{juraszek2006sampling}. The Amber03 force field employed in this work possesses an elevated propensity for $\alpha$-helicity relative to the OPLSAA force field employed by Juraszek \textit{et al.}\ \cite{best2008current,juraszek2006sampling,paschek2008computing}, which we suggest may be at the root of the high stability of the $\alpha$-helix observed in our simulations.

Similarly, our FES also shows good correspondence with that determined by Kim \textit{et al.}\ \cite{kim2015systematic} employing the Amber ff03w force field\cite{best2010protein} and TIP4P/2005 water model\cite{abascal2005general}. In particular, the line of metastable basins at helix-RMSD $\approx$ 0.3 \AA\ map to the \textbf{F}, \textbf{I}, \textbf{J}, and \textbf{K} basins in the vicinity of the native fold identified in that work. However, our landscape only extends up to helix-RMSD values of $\sim$2 \AA\ with the N-terminal $\alpha$-helix always intact, whereas Kim \textit{et al.}\ resolve up to helix-RMSD values of 4.5 \AA. These authors achieve helix unfolding by high-temperature denaturation and estimate the FES from 25 independent folding trajectories that are terminated upon attaining the native fold. Accordingly -- as the authors observe -- the simulation data are not sampled from an equilibrated distribution, and the FES cannot be considered to be thermodynamically converged. Accordingly, we suggest that the high-helix-RMSD configurations resolved by Kim \textit{et al.}\ actually lie far higher in free energy, are not thermally accessible at room temperature, and are \blauwr{a result} of the simulation protocol in which folding runs were initialized from thermally denatured starting structures. The folding pathway resolved by MESA shows good correspondence with the latter portion of the diffusion-collision pathway (Path B) identified by Kim \textit{et al.}\ using diffusion maps. We also resolve the metastable loop structure \textbf{L} termed \textbf{LOOP-I} by Kim \textit{et al.}, the near-native intermediate \textbf{B} that is termed \textbf{HLX-I}. We do not resolve the upper portion of the diffusion-collision pathway, nor the nucleation-condensation pathway (Path A), both of which correspond to folding of the initially unfolded N-terminal $\alpha$-helix. Interestingly, the structural details and relative propensity of the nucleation-condensation pathways identified by Juraszek \textit{et al.}\ and Kim \textit{et al.}\ are not in agreement. Kim \textit{et al.}\ ascribe these discrepancies to differences in the force field, but it is also likely a result of the nonequilibrium manner in which data were collected in that work \cite{kim2015systematic}. We also suggest that the root of the approximate symmetry in CV $\phi_2$ in the horseshoe-like landscape resolved by Kim \textit{et al.} is the global molecular chirality \cite{kim2015systematic} -- which is the same origin of our approximate symmetry in CV $\xi_1$ -- and that this CV does not distinguish the two folding mechanisms. Careful inspection of Fig.~8 in that work in which points are colored by helix-RMSD, suggests that both the diffusion-collision and nucleation condensation pathways can proceed under each global chirality \cite{kim2015systematic}. In our work, we also observe the diffusion-collision path to proceed along the left- and right-handed chiralities, but do not resolve the nucleation-condensation routes since the N-terminal $\alpha$-helix remains intact.

Our folding pathways show the best correspondence with those identified by Paschek \textit{et al.}\ \cite{paschek2008computing}, who employed REMD simulations using the Amber ff94 force field\cite{cornell1995second} and TIP3P water model \cite{jorgensen1983comparison}. In excellent agreement with our calculations, these authors identified a single diffusion-collision folding pathway in good correspondence with that identified by both Juraszek \textit{et al.}\ \cite{juraszek2006sampling} and Kim \textit{et al.}\ \cite{kim2015systematic}. Paschek \textit{et al.}\ share our concerns, however, that the excessive stability of helices in the protein force field may artificially overemphasize this folding route \cite{paschek2008computing}. Our results are also in good agreement with the work of Deng \textit{at al.}\ \cite{deng2013kinetics}, who analyzed using Markov state models a long 208 $\mu$s simulation performed by Lindorff-Larsen \textit{et al.}\cite{lindorff2011fast}  employing the Charm22 force field \cite{mackerell1998all,mackerell2004extending} and TIP3P water model \cite{jorgensen1983comparison} These researchers also resolved competing diffusion-collision and nucleation-condensation pathways, but found the former to be overwhelmingly more probable than the latter.

In sum, MESA achieves automated data-driven discovery of two CVs parameterizing the important collective motions of Trp-cage and in which it executes accelerated sampling of configurational space. The resultant FES in these two CVs identifies the important (meta)stable configurational states, and our accelerated sampling protocol probes a $\sim$20 $k_B T$ range in free energy. The FES, and folding pathways it contains, are in good agreement with prior work \cite{juraszek2006sampling,kim2015systematic,paschek2008computing,deng2013kinetics}, although in the present study we do not observe unfolding of the N-terminal $\alpha$-helix and therefore do not resolve the nucleation-condensation folding pathway. The stability of the N-terminal $\alpha$-helix is likely largely attributable to the favorability of helical secondary structure within the Amber03 force field \cite{wang2004development,best2008current,paschek2008computing}, and it would be of interest to compare the CVs, FES, and folding pathways resolved by MESA under different choices of protein force fields and solvent models. 

We observe that MESA also offers a means to enhance configurational sampling of particular secondary structural elements by tuning the error function used to train the autoencoder to focus on particular regions of the protein (\blauw{Eqn.~\ref{err_3}}). Our current choice of error function trains the network to optimally reconstruct -- under translation and rigid rotation -- the protein coordinates, wherein the contribution of each atom to the reconstruction error is equally weighted. By adapting the error function to more heavily weight particular atoms, we can train the network to focus on accurate reconstruction of particular regions of the protein secondary structure. In the present case, placing more weight on the atoms residing within the N-terminal $\alpha$-helix may nudge MESA to discover CVs tailored to this region and assist in improved sampling of secondary structural changes in this region.

\begin{figure}[ht!]
\begin{center}
\includegraphics[width=.50\textwidth]{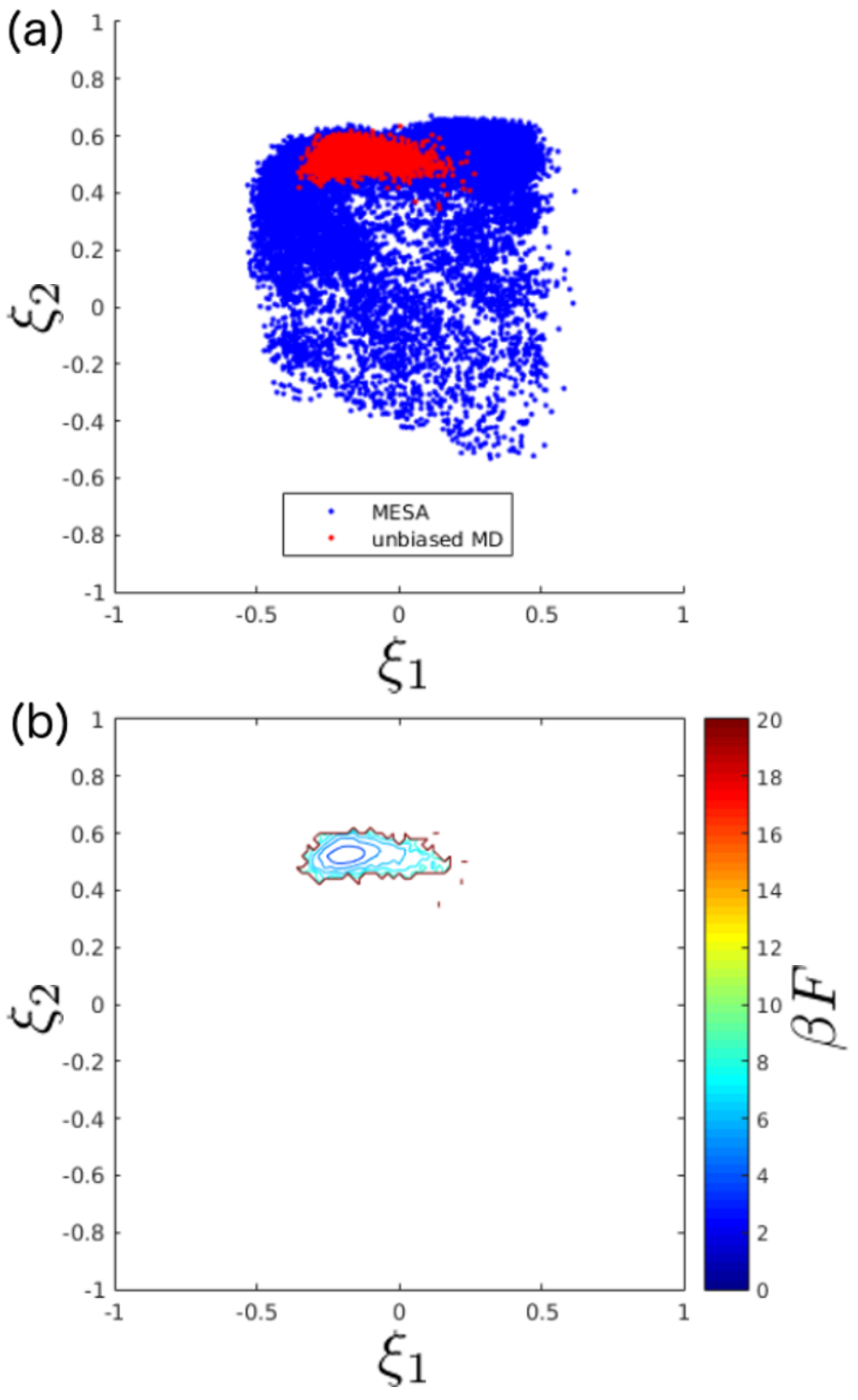} 
\caption{Benchmarking MESA acceleration for Trp-cage. (a) Projection into the MESA identified CVs $\{\xi_1,\xi_2\}$ the collated configurations collected throughout the six iterations of MESA (blue), and a 300 ns unbiased MD simulation commencing from the Trp-cage native state (red). Both MESA and the unbiased run require 72 GPU-hours of execution time. (b) The free energy surface $F(\xi_1,\xi_2)$ computed from the unbiased simulation trajectory resolves only the native basin and samples a $\sim$7 $k_B T$ free energy range, compared to the far more comprehensive configurational sampling and $\sim$20 $k_B T$ range achieved by MESA (\blauw{Fig.~\ref{FES_Trp_cage}a}).}
\label{Trp_bench}
\end{center}
\end{figure}

\textbf{Benchmarking of MESA acceleration.} In aggregate, the six rounds of MESA required 72 GPU-hours on a GeForce GTX 960 card, with network training costs dominated by the $\sim$300 ns of molecular simulation. To quantify the acceleration in exploration of configurational space offered by MESA, we conducted a 72 GPU-hour unbiased simulation commencing from the Trp-cage native state to generate a 300 ns simulation trajectory. Projection in \blauw{Fig.~\ref{Trp_bench}a} of the simulation snapshots resulting from this unbiased calculation into the $\{\xi_1,\xi_2\}$ CVs of the terminal autoencoder reveals that it is entirely confined to the native basin, and samples a tiny fraction of the configurational space explored by MESA. The corresponding unbiased FES in \blauw{Fig.~\ref{Trp_bench}b} shows that the unbiased run probes a free energy range of only $\sim$7 $k_B T$, compared to the far more expansive sampling and $\sim$20 $k_B T$ range explored by MESA (\blauw{Fig.~\ref{FES_Trp_cage}a}).

\section{\sffamily \Large Conclusions} \label{sec:concl}

In this work, we have introduced a new methodology -- \underline{M}olecular \underline{E}nhanced \underline{S}ampling with \underline{A}utoencoders (\groen{MESA}) -- for the data-driven accelerated order parameter discovery and accelerated sampling of the configurational space of macromolecular and biomolecular systems. The methodology requires minimal prior knowledge of the system, requiring only an initial short simulation trajectory to seed iterative, on-the-fly discovery of low-dimensional collective variables and accelerated sampling of configurational space. The essence of our approach is to iteratively train auto-associative artificial neural networks (autoencoders) to learn low-dimensional projections into nonlinear collective variables (\groen{CVs}) coincident with the important collective molecular motions and which -- in contrast with other nonlinear manifold learning approaches -- are explicit and differentiable functions of the atomic coordinates. The availability of this explicit mapping admits analytical expressions propagating biasing potentials in the CVs to biasing forces on the atoms, thereby enabling accelerated sampling directly in the data-driven CVs using collective variable biasing techniques. The capacity to apply biasing forces directly within the nonlinear CVs distinguishes our approach from existing data-driven accelerated sampling techniques that appeal to proxy variables, basis set projections, local linear dimensionality reduction, or judicious initialization of unbiased calculations to enhance sampling. We perform interleaved rounds of autoencoder CV discovery and biased sampling until we achieve convergence in the CVs and region of explored configurational space. We have largely automated the MESA framework to require minimal user intervention, and made it freely available for public download (\url{https://github.com/weiHelloWorld/accelerated_sampling_with_autoencoder}) together with a plugin implementing the biasing forces within the OpenMM simulation package \cite{eastman2017openmm,friedrichs2009accelerating} (\url{https://github.com/weiHelloWorld/ANN_Force}).

We have validated MESA in simulations of alanine dipeptide in vacuum and Trp-cage in explicit water. In each case we recover low-dimensional nonlinear CVs and efficiently drive sampling of configurational space to recover converged free energy surfaces (\groen{FES}) resolving the important metastable states and folding pathways. Our application to alanine dipeptide illuminates the difficulties associated with periodicities in the inherent manifold -- in this case a flat torus -- and we demonstrate how to confront this issue by increasing the dimensionality of the embedding space to fully unfold the manifold and prevent spurious intersections. In the case of Trp-cage, we recover a FES and diffusion-collision folding mechanism in good agreement with prior work. The favorability of helical secondary structure within the protein force field means that we did not observe unfolding of the N-terminal $\alpha$-helix, and therefore did not resolve the nucleation-condensation folding pathway. It would be of interest in future work to compare the MESA CVs and FES predictions under different choices of protein and solvent potential functions. Benchmarking of the acceleration offered by MESA in sampling the Trp-cage configurational space demonstrated a three-fold increase in the sampled range of free energy and massive enhancement over the exploration achieved by unbiased calculations utilizing the same computational resources. We expect MESA to prove most valuable for large biomolecular systems about which there is little expert knowledge of the important conformational motions and where expensive computation is at a premium. \blauwr{Preliminary testing demonstrates that autoencoders with 15,000-1000-3-1000-15,000 topologies can be efficiently trained over 32,000 configurations in under 40 GPU-minutes, supporting the scalability of network training to large molecular systems. Generation of sufficient quantities of training data through biased simulation and robust application of regularization are of prime concern in achieving good network performance while avoiding overfitting in these sizable networks. We note that the computational burden of network training may be alleviated through coarse-grained representations of the molecule to the autoencoder employing, for example, only backbone atom coordinates or internal dihedral angles, and of biased simulation through the use of coarse-grained or multi-resolution molecular models \cite{kmiecik2016coarse}.}

We see many avenues for future development of our approach. First, we have employed umbrella sampling as an efficient and embarrassingly parallel accelerated sampling approach, but anticipate that variants of metadynamics \cite{laio2002escaping,huber1994local,barducci2008well} may offer an attractive alternative by eliminating the need for boundary detection, initialization of swarms of independent umbrella runs, and adaptive tuning of umbrella potential force constants. Second, the development of bespoke error functions used to train the network presents a powerful means to incorporate prior system knowledge and focus the calculated CVs and accelerated sampling on particular regions and features of the system. For example, this approach may be of value in unfolding particular secondary structure elements, focusing sampling on enzymatic active sites, or actuating important allosteric motions. One drawback of the current error function is that there is no ranking of the CVs discovered by the autoencoder, such that the relative importance of the collective modes is not apparent. The hierarchical error functions proposed by Scholz and Vig\'{a}rio \cite{scholz2008nonlinear,scholz2002nonlinear} extract CVs within a nested hierarchy of subnetworks that discover successively higher dimensional projections that minimize the reconstruction error subject to the constraint that all lower dimensional projections also minimize the reconstruction error. This imposition of a rank ordering on the CVs can help interpret the relative importance of the identified collective motions and also stabilize the CV discovery \cite{scholz2008nonlinear}. Third, rather than training a single network over all of the collated simulation data in each iteration to discover and refine good global CVs, we propose that it may be beneficial to train independent networks over different regions of the inherent manifold in order to discover good local CVs. In the spirit of local tangent space alignment \cite{wang2011geometric} and iMapD \cite{chiavazzo2017intrinsic}, local CVs adapted to their neighborhood of configurational space can resolve the collective motions relevant to that region and improve the efficiency of the enhanced sampling. The biased sampling data furnished from each network may then be patched together using reweighting techniques \cite{ferguson2017bayeswham}, and the terminal landscape constructed in the union of all local CVs or a unified set of global CVs identified \textit{post hoc} in the final MESA iteration.

\section*{\sffamily \Large Acknowledgments}

This material is based upon work supported by the National Science Foundation under Grant No.~CHE-1664426.


\clearpage
\newpage

\section*{\sffamily \Large Appendix}

\setcounter{figure}{0}
\renewcommand{\thefigure}{A\arabic{figure}}

\begin{figure}[ht!]
\begin{center}
\includegraphics[width=0.82\textwidth]{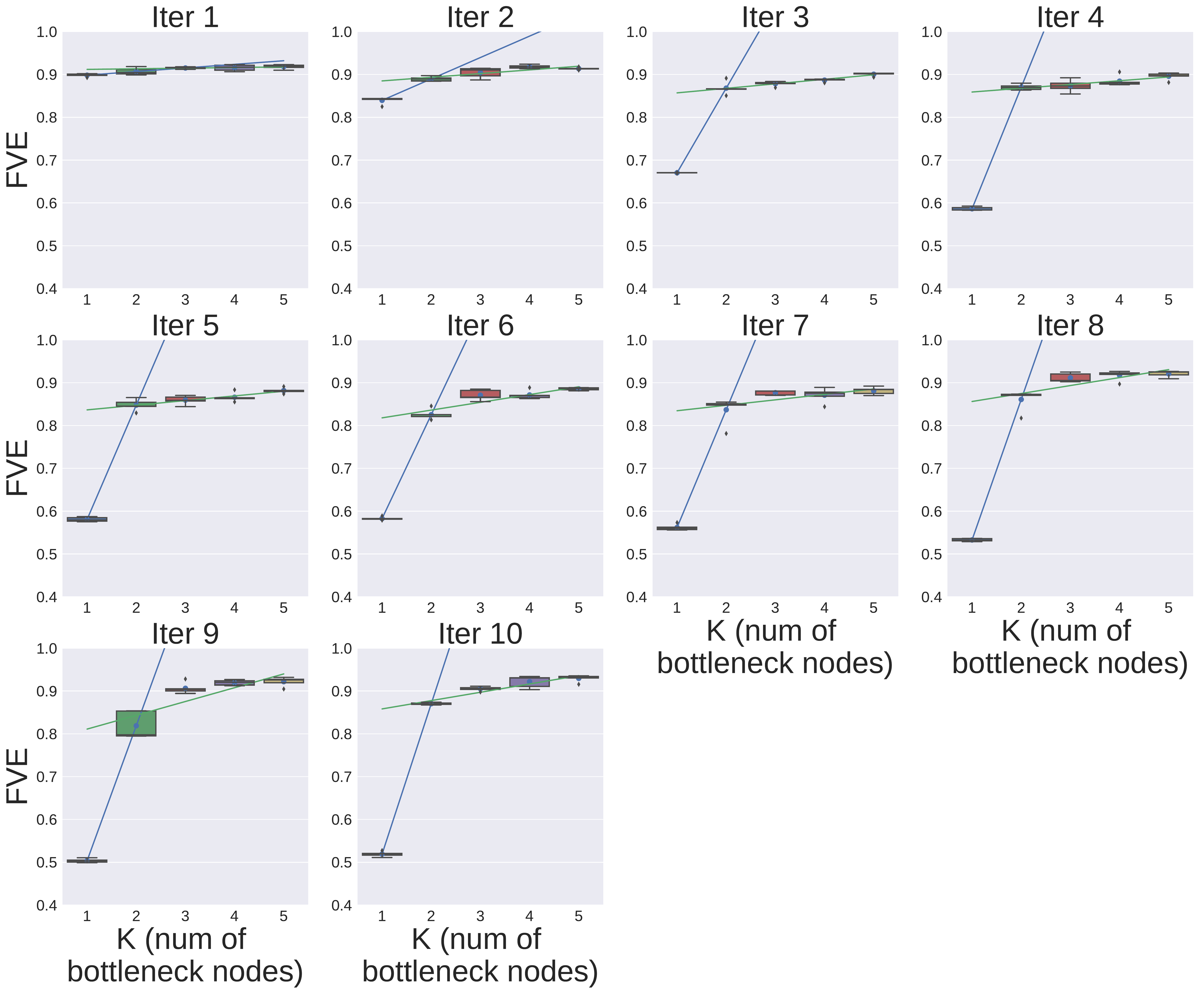} 
\caption{Determination of the dimensionality of the intrinsic manifold in each iteration of MESA applied to alanine dipeptide in vacuum. The fraction of variance explained (FVE) by the autoencoder reconstructed outputs as a function of the number of bottleneck nodes $K$ was computed according to \blauw{Eqn.~\ref{FVE}}. For each value of $K$ at each iteration, we independently initialized and trained five autoencoders with identical topologies over the same data and plotted the calculated distribution as a boxplot. The boxes span the upper and lower quartiles of the FVE distribution, while the whiskers show the range of the distribution with the exception of outliers that are rendered as points. A knee in each plot was ascertained using the L-method that determines the two piecewise linear fits that best explain the data \cite{salvador2004determining}. The dimensionality of the intrinsic manifold is determined to be $K$=2 for all 10 iterations of MESA. Boxplots were generated using the seaborn Python libraries (\url{https://seaborn.pydata.org}).}
\label{L_dim_Ala}
\end{center}
\end{figure}

\begin{figure}[ht!]
\begin{center}
\includegraphics[width=0.60\textwidth]{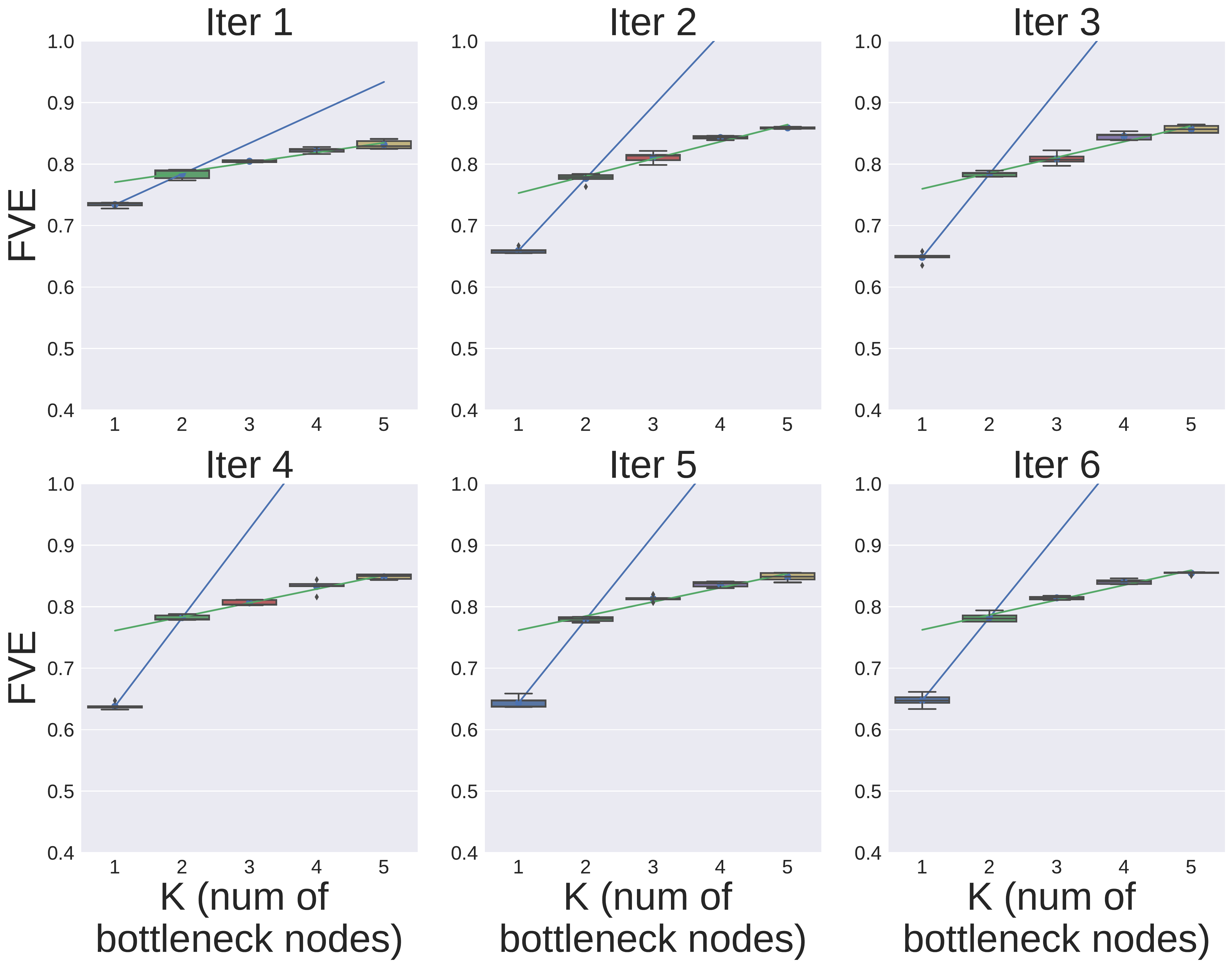} 
\caption{Determination of the dimensionality of the intrinsic manifold in each iteration of MESA applied to Trp-cage in water. Construction, interpretation, and analysis of the boxplots is identical to that in \blauw{Fig.~\ref{L_dim_Ala}}. The dimensionality of the intrinsic manifold is determined to be $K$=2 for all 6 iterations of MESA.}
\label{L_dim_Trp_cage}
\end{center}
\end{figure}


\clearpage
\newpage

\bibliography{library}

\begin{thebibliography}{100}

\bibitem{hashemian2013modeling}
Hashemian, B.; Mill{\'a}n, D. and Arroyo, M., {\em The Journal of Chemical
  Physics}, 2013, {\bf 139}(21).

\bibitem{bernardi2015enhanced}
Bernardi, R.~C.; Melo, M.~C. and Schulten, K., {\em Biochimica et Biophysica
  Acta}, 2015, {\bf 1850}(5), 872--877.

\bibitem{karplus1990molecular}
Karplus, M. and Petsko, G.~A., {\em Nature}, 1990, {\bf 347}(6294), 631--639.

\bibitem{rohrdanz2013discovering}
Rohrdanz, M.~A.; Zheng, W. and Clementi, C., {\em Annual Review of Physical
  Chemistry}, 2013, {\bf 64}, 295--316.

\bibitem{abrams2013enhanced}
Abrams, C. and Bussi, G., {\em Entropy}, 2013, {\bf 16}(1), 163--199.

\bibitem{dellago1998transition}
Dellago, C.; Bolhuis, P.~G.; Csajka, F.~S. and Chandler, D., {\em The Journal
  of Chemical Physics}, 1998, {\bf 108}(5), 1964--1977.

\bibitem{bolhuis2002transition}
Bolhuis, P.~G.; Chandler, D.; Dellago, C. and Geissler, P.~L., {\em Annual
  Review of Physical Chemistry}, 2002, {\bf 53}(1), 291--318.

\bibitem{rogal2008multiple}
Rogal, J. and Bolhuis, P.~G., {\em The Journal of Chemical Physics}, 2008, {\bf
  129}(22), 224107.

\bibitem{moroni2004rate}
Moroni, D.; Bolhuis, P.~G. and van Erp, T.~S., {\em The Journal of Chemical
  Physics}, 2004, {\bf 120}(9), 4055--4065.

\bibitem{van2003novel}
van Erp, T.~S.; Moroni, D. and Bolhuis, P.~G., {\em The Journal of Chemical
  Physics}, 2003, {\bf 118}(17), 7762--7774.

\bibitem{weinan2002string}
Weinan, E.; Ren, W. and Vanden-Eijnden, E., {\em Physical Review B}, 2002, {\bf
  66}(5), 052301.

\bibitem{weinan2005finite}
Weinan, E.; Ren, W. and Vanden-Eijnden, E., {\em Journal of Physical Chemistry
  B}, 2005, {\bf 109}(14), 6688--6693.

\bibitem{jonsson1998nudged}
J{\'o}nsson, H.; Mills, G. and Jacobsen, K.~W. In {\em Classical and Quantum
  Dynamics in Condensed Phase Simulations}, Berne, B.~J.; Ciccotti, G. and
  Coker, D.~F., Eds.;
\newblock World Scientific, Singapore, 1998;
\newblock page 385.

\bibitem{sheppard2008optimization}
Sheppard, D.; Terrell, R. and Henkelman, G., {\em The Journal of Chemical
  Physics}, 2008, {\bf 128}(13), 134106.

\bibitem{borrero2007reaction}
Borrero, E.~E. and Escobedo, F.~A., {\em The Journal of Chemical Physics},
  2007, {\bf 127}(16), 164101.

\bibitem{allen2009forward}
Allen, R.~J.; Valeriani, C. and ten Wolde, P.~R., {\em Journal of Physics:
  Condensed Matter}, 2009, {\bf 21}(46), 463102.

\bibitem{brooks1995optimization}
Brooks, S.~P. and Morgan, B.~J., {\em The Statistician}, 1995, pages 241--257.

\bibitem{berg1992multicanonical}
Berg, B.~A. and Neuhaus, T., {\em Physical Review Letters}, 1992, {\bf 68}(1),
  9.

\bibitem{hansmann1997parallel}
Hansmann, U.~H., {\em Chemical Physics Letters}, 1997, {\bf 281}(1), 140--150.

\bibitem{sugita1999replica}
Sugita, Y. and Okamoto, Y., {\em Chemical Physics Letters}, 1999, {\bf 314}(1),
  141--151.

\bibitem{sugita2000replica}
Sugita, Y. and Okamoto, Y., {\em Chemical Physics Letters}, 2000, {\bf 329}(3),
  261--270.

\bibitem{mitsutake2001generalized}
Mitsutake, A.; Sugita, Y. and Okamoto, Y., {\em Peptide Science}, 2001, {\bf
  60}(2), 96--123.

\bibitem{fukunishi2002hamiltonian}
Fukunishi, H.; Watanabe, O. and Takada, S., {\em The Journal of Chemical
  Physics}, 2002, {\bf 116}(20), 9058--9067.

\bibitem{okamoto2004generalized}
Okamoto, Y., {\em Journal of Molecular Graphics and Modelling}, 2004, {\bf
  22}(5), 425--439.

\bibitem{liu2006hydrophobic}
Liu, P.; Huang, X.; Zhou, R. and Berne, B.~J., {\em The Journal of Physical
  Chemistry B}, 2006, {\bf 110}(38), 19018--19022.

\bibitem{torrie1977nonphysical}
Torrie, G.~M. and Valleau, J.~P., {\em Journal of Computational Physics}, 1977,
  {\bf 23}(2), 187--199.

\bibitem{voter1997hyperdynamics}
Voter, A.~F., {\em Physical Review Letters}, 1997, {\bf 78}(20), 3908.

\bibitem{grubmuller1995predicting}
Grubm{\"u}ller, H., {\em Physical Review E}, 1995, {\bf 52}(3), 2893.

\bibitem{laio2002escaping}
Laio, A. and Parrinello, M., {\em Proceedings of the National Academy of
  Sciences}, 2002, {\bf 99}(20), 12562--12566.

\bibitem{huber1994local}
Huber, T.; Torda, A.~E. and van Gunsteren, W.~F., {\em Journal of
  Computer-Aided Molecular Design}, 1994, {\bf 8}(6), 695--708.

\bibitem{barducci2008well}
Barducci, A.; Bussi, G. and Parrinello, M., {\em Physical Review Letters},
  2008, {\bf 100}(2), 020603.

\bibitem{rosso2002use}
Rosso, L.; Min{\'a}ry, P.; Zhu, Z. and Tuckerman, M.~E., {\em The Journal of
  Chemical Physics}, 2002, {\bf 116}(11), 4389--4402.

\bibitem{maragliano2006temperature}
Maragliano, L. and Vanden-Eijnden, E., {\em Chemical Physics Letters}, 2006,
  {\bf 426}(1), 168--175.

\bibitem{abrams2008efficient}
Abrams, J.~B. and Tuckerman, M.~E., {\em The Journal of Physical Chemistry B},
  2008, {\bf 112}(49), 15742--15757.

\bibitem{so2000temperature}
So/rensen, M.~R. and Voter, A.~F., {\em The Journal of Chemical Physics}, 2000,
  {\bf 112}(21), 9599--9606.

\bibitem{den1998calculation}
den Otter, W.~K. and Briels, W.~J., {\em The Journal of Chemical Physics},
  1998, {\bf 109}(11), 4139--4146.

\bibitem{carter1989constrained}
Carter, E.; Ciccotti, G.; Hynes, J.~T. and Kapral, R., {\em Chemical Physics
  Letters}, 1989, {\bf 156}(5), 472--477.

\bibitem{ciccotti2004blue}
Ciccotti, G. and Ferrario, M., {\em Molecular Simulation}, 2004, {\bf
  30}(11-12), 787--793.

\bibitem{darve2008adaptive}
Darve, E.; Rodr{\'\i}guez-G{\'o}mez, D. and Pohorille, A., {\em The Journal of
  Chemical Physics}, 2008, {\bf 128}(14), 144120.

\bibitem{kirkwood1935statistical}
Kirkwood, J.~G., {\em The Journal of Chemical Physics}, 1935, {\bf 3}(5),
  300--313.

\bibitem{straatsma1988free}
Straatsma, T. and Berendsen, H., {\em The Journal of Chemical Physics}, 1988,
  {\bf 89}(9), 5876--5886.

\bibitem{wang2017nonlinear}
Wang, J. and Ferguson, A., {\em Molecular Simulation}, 2017, pages 1--18.

\bibitem{hegger2007complex}
Hegger, R.; Altis, A.; Nguyen, P.~H. and Stock, G., {\em Physical Review
  Letters}, 2007, {\bf 98}(2), 028102.

\bibitem{ferguson2011nonlinear}
Ferguson, A.~L.; Panagiotopoulos, A.~Z.; Kevrekidis, I.~G. and Debenedetti,
  P.~G., {\em Chemical Physics Letters}, 2011, {\bf 509}(1), 1--11.

\bibitem{ferguson2010systematic}
Ferguson, A.~L.; Panagiotopoulos, A.~Z.; Debenedetti, P.~G. and Kevrekidis,
  I.~G., {\em Proceedings of the National Academy of Sciences}, 2010, {\bf
  107}(31), 13597--13602.

\bibitem{zhuravlev2009deconstructing}
Zhuravlev, P.~I.; Materese, C.~K. and Papoian, G.~A., {\em The Journal of
  Physical Chemistry B}, 2009, {\bf 113}(26), 8800--8812.

\bibitem{amadei1993essential}
Amadei, A.; Linssen, A. and Berendsen, H.~J., {\em Proteins: Structure,
  Function, and Bioinformatics}, 1993, {\bf 17}(4), 412--425.

\bibitem{garcia1992large}
Garc{\'\i}a, A.~E., {\em Physical Review Letters}, 1992, {\bf 68}(17), 2696.

\bibitem{das2006low}
Das, P.; Moll, M.; Stamati, H.; Kavraki, L.~E. and Clementi, C., {\em
  Proceedings of the National Academy of Sciences}, 2006, {\bf 103}(26),
  9885--9890.

\bibitem{stamati2010application}
Stamati, H.; Clementi, C. and Kavraki, L.~E., {\em Proteins: Structure,
  Function, and Bioinformatics}, 2010, {\bf 78}(2), 223--235.

\bibitem{ichiye1991collective}
Ichiye, T. and Karplus, M., {\em Proteins: Structure, Function, and
  Bioinformatics}, 1991, {\bf 11}(3), 205--217.

\bibitem{pearson1901liii}
Pearson, K., {\em The London, Edinburgh, and Dublin Philosophical Magazine and
  Journal of Science}, 1901, {\bf 2}(11), 559--572.

\bibitem{troyer1995protein}
Troyer, J.~M. and Cohen, F.~E., {\em Proteins: Structure, Function, and
  Bioinformatics}, 1995, {\bf 23}(1), 97--110.

\bibitem{scholkopf1997kernel}
Sch{\"o}lkopf, B.; Smola, A. and M{\"u}ller, K.-R. In {\em International
  Conference on Artificial Neural Networks}, pages 583--588. Springer, 1997.

\bibitem{wang2011geometric}
Wang, J., {\em Geometric Structure of High-Dimensional Data and Dimensionality
  Reduction}, Springer, 2011.

\bibitem{roweis2000nonlinear}
Roweis, S.~T. and Saul, L.~K., {\em Science}, 2000, {\bf 290}(5500),
  2323--2326.

\bibitem{zhang2007mlle}
Zhang, Z. and Wang, J. In {\em Advances in Neural Information Processing
  Systems}, pages 1593--1600, 2007.

\bibitem{tenenbaum2000global}
Tenenbaum, J.~B.; De~Silva, V. and Langford, J.~C., {\em Science}, 2000, {\bf
  290}(5500), 2319--2323.

\bibitem{weinberger2006unsupervised}
Weinberger, K.~Q. and Saul, L.~K., {\em International Journal of Computer
  Vision}, 2006, {\bf 70}(1), 77--90.

\bibitem{li2006version}
Li, C.-g.; Guo, J.; Chen, G.; Nie, X.-f. and Yang, Z. In {\em 2006
  International Conference on Machine Learning and Cybernetics}, pages
  3201--3206. IEEE, 2006.

\bibitem{belkin2002laplacian}
Belkin, M. and Niyogi, P. In {\em Advances in Neural Information Processing
  Systems}, pages 585--591, 2002.

\bibitem{donoho2003hessian}
Donoho, D.~L. and Grimes, C., {\em Proceedings of the National Academy of
  Sciences}, 2003, {\bf 100}(10), 5591--5596.

\bibitem{ferguson2011integrating}
Ferguson, A.~L.; Panagiotopoulos, A.~Z.; Debenedetti, P.~G. and Kevrekidis,
  I.~G., {\em The Journal of Chemical Physics}, 2011, {\bf 134}(13), 135103.

\bibitem{coifman2006diffusion}
Coifman, R.~R. and Lafon, S., {\em Applied and Computational Harmonic
  Analysis}, 2006, {\bf 21}(1), 5--30.

\bibitem{rohrdanz2011determination}
Rohrdanz, M.~A.; Zheng, W.; Maggioni, M. and Clementi, C., {\em The Journal of
  Chemical Physics}, 2011, {\bf 134}(12), 03B624.

\bibitem{preto2014fast}
Preto, J. and Clementi, C., {\em Physical Chemistry Chemical Physics}, 2014,
  {\bf 16}(36), 19181--19191.

\bibitem{ceriotti2011simplifying}
Ceriotti, M.; Tribello, G.~A. and Parrinello, M., {\em Proceedings of the
  National Academy of Sciences}, 2011, {\bf 108}(32), 13023--13028.

\bibitem{tribello2012using}
Tribello, G.~A.; Ceriotti, M. and Parrinello, M., {\em Proceedings of the
  National Academy of Sciences}, 2012, {\bf 109}(14), 5196--5201.

\bibitem{ceriotti2013demonstrating}
Ceriotti, M.; Tribello, G.~A. and Parrinello, M., {\em Journal of Chemical
  Theory and Computation}, 2013, {\bf 9}(3), 1521--1532.

\bibitem{ferguson2010experimental}
Ferguson, A.~L.; Zhang, S.; Dikiy, I.; Panagiotopoulos, A.~Z.; Debenedetti,
  P.~G. and Link, A.~J., {\em Biophysical Journal}, 2010, {\bf 99}(9),
  3056--3065.

\bibitem{zheng2011polymer}
Zheng, W.; Rohrdanz, M.~A.; Maggioni, M. and Clementi, C., {\em The Journal of
  Chemical Physics}, 2011, {\bf 134}(14), 144109.

\bibitem{zheng2013rapid}
Zheng, W.; Rohrdanz, M.~A. and Clementi, C., {\em The journal of physical
  chemistry B}, 2013, {\bf 117}(42), 12769--12776.

\bibitem{spiwok2011metadynamics}
Spiwok, V. and Kr{\'a}lov{\'a}, B., {\em The Journal of Chemical Physics},
  2011, {\bf 135}(22), 224504.

\bibitem{fiorin2013using}
Fiorin, G.; Klein, M.~L. and H{\'e}nin, J., {\em Molecular Physics}, 2013, {\bf
  111}(22-23), 3345--3362.

\bibitem{ma2005automatic}
Ma, A. and Dinner, A.~R., {\em The Journal of Physical Chemistry B}, 2005, {\bf
  109}(14), 6769--6779.

\bibitem{peters2006obtaining}
Peters, B. and Trout, B.~L., {\em The Journal of Chemical Physics}, 2006, {\bf
  125}(5), 054108.

\bibitem{peters2007extensions}
Peters, B.; Beckham, G.~T. and Trout, B.~L., {\em The Journal of Chemical
  Physics}, 2007, {\bf 127}(3), 034109.

\bibitem{abrams2012fly}
Abrams, C.~F. and Vanden-Eijnden, E., {\em Chemical Physics Letters}, 2012,
  {\bf 547}, 114--119.

\bibitem{branduardi2007b}
Branduardi, D.; Gervasio, F.~L. and Parrinello, M., {\em The Journal of
  Chemical Physics}, 2007, {\bf 126}(5), 054103.

\bibitem{chiavazzo2017intrinsic}
Chiavazzo, E.; Covino, R.; Coifman, R.~R.; Gear, C.~W.; Georgiou, A.~S.;
  Hummer, G. and Kevrekidis, I.~G., {\em Proceedings of the National Academy of
  Sciences}, 2017, page 201621481.

\bibitem{scholz2002nonlinear}
Scholz, M. and Vig{\'a}rio, R. In {\em ESANN'2002 proceedings - European
  Symposium on Artificial Neural Networks}, pages 439--444, 2002.

\bibitem{scholz2008nonlinear}
Scholz, M.; Fraunholz, M. and Selbig, J. In {\em Principal manifolds for data
  visualization and dimension reduction;}
\newblock Springer, 2008;
\newblock pages 44--67.

\bibitem{hinton2006reducing}
Hinton, G.~E. and Salakhutdinov, R.~R., {\em Science}, 2006, {\bf 313}(5786),
  504--507.

\bibitem{yan2007graph}
Yan, S.; Xu, D.; Zhang, B.; Zhang, H.-J.; Yang, Q. and Lin, S., {\em IEEE
  Transactions on Pattern Analysis and Machine Intelligence}, 2007, {\bf
  29}(1), 40--51.

\bibitem{wang2014generalized}
Wang, W.; Huang, Y.; Wang, Y. and Wang, L. In {\em Proceedings of the IEEE
  Conference on Computer Vision and Pattern Recognition Workshops}, pages
  490--497, 2014.

\bibitem{eastman2017openmm}
Eastman, P.; Swails, J.; Chodera, J.~D.; McGibbon, R.~T.; Zhao, Y.; Beauchamp,
  K.~A.; Wang, L.-P.; Simmonett, A.~C.; Harrigan, M.~P. and Stern, C.~D., {\em
  PLOS Computational Biology}, 2017, {\bf 13}(7), e1005659.

\bibitem{eastman2012openmm}
Eastman, P.; Friedrichs, M.~S.; Chodera, J.~D.; Radmer, R.~J.; Bruns, C.~M.;
  Ku, J.~P.; Beauchamp, K.~A.; Lane, T.~J.; Wang, L.-P. and Shukla, D., {\em
  Journal of Chemical Theory and Computation}, 2012, {\bf 9}(1), 461--469.

\bibitem{friedrichs2009accelerating}
Friedrichs, M.~S.; Eastman, P.; Vaidyanathan, V.; Houston, M.; Legrand, S.;
  Beberg, A.~L.; Ensign, D.~L.; Bruns, C.~M. and Pande, V.~S., {\em Journal of
  Computational Chemistry}, 2009, {\bf 30}(6), 864--872.

\bibitem{hassoun1995fundamentals}
Hassoun, M.~H., {\em Fundamentals of Artificial Neural Networks}, MIT press,
  1995.

\bibitem{mcculloch1943logical}
McCulloch, W.~S. and Pitts, W., {\em The Bulletin of Mathematical Biophysics},
  1943, {\bf 5}(4), 115--133.

\bibitem{chen1995universal}
Chen, T. and Chen, H., {\em IEEE Transactions on Neural Networks}, 1995, {\bf
  6}(4), 911--917.

\bibitem{kramer1991nonlinear}
Kramer, M.~A., {\em AIChE Journal}, 1991, {\bf 37}(2), 233--243.

\bibitem{kirby1996circular}
Kirby, M.~J. and Miranda, R., {\em Neural Computation}, 1996, {\bf 8}(2),
  390--402.

\bibitem{bourlard1988auto}
Bourlard, H. and Kamp, Y., {\em Biological Cybernetics}, 1988, {\bf 59}(4),
  291--294.

\bibitem{friedman2001elements}
Friedman, J.; Hastie, T. and Tibshirani, R., {\em The Elements of Statistical
  Learning}, Vol. ~1, Springer series in statistics New York, 2001.

\bibitem{collobert2004links}
Collobert, R. and Bengio, S. In {\em Proceedings of the twenty-first
  international conference on Machine learning}, page~23. ACM, 2004.

\bibitem{bengio2012practical}
Bengio, Y. In {\em Neural networks: Tricks of the trade;}
\newblock Springer, 2012;
\newblock pages 437--478.

\bibitem{rumelhart1995backpropagation}
Rumelhart, D.~E.; Durbin, R.; Golden, R. and Chauvin, Y., {\em Backpropagation:
  Theory, architectures and applications}, 1995, pages 1--34.

\bibitem{rumelhart1988learning}
Rumelhart, D.~E.; Hinton, G.~E. and Williams, R.~J. In {\em Cognitive
  Modeling}, Polk, T.~A. and Seifert, C.~M., Eds.;
\newblock MIT Press, 1988;
\newblock chapter~8.

\bibitem{sutskever2013importance}
Sutskever, I.; Martens, J.; Dahl, G. and Hinton, G. In {\em International
  conference on machine learning}, pages 1139--1147, 2013.

\bibitem{salvador2004determining}
Salvador, S. and Chan, P. In {\em Tools with Artificial Intelligence, 2004.
  ICTAI 2004. 16th IEEE International Conference on}, pages 576--584. IEEE,
  2004.

\bibitem{krizhevsky2012imagenet}
Krizhevsky, A.; Sutskever, I. and Hinton, G.~E. In {\em Advances in Neural
  Information Processing Systems}, pages 1097--1105, 2012.

\bibitem{kabsch1976solution}
Kabsch, W., {\em Acta Crystallographica Section A: Crystal Physics,
  Diffraction, Theoretical and General Crystallography}, 1976, {\bf 32}(5),
  922--923.

\bibitem{roux1995calculation}
Roux, B., {\em Computer Physics Communications}, 1995, {\bf 91}(1-3), 275--282.

\bibitem{chipot2007free}
Chipot, C. and Pohorille, A., {\em Free Energy Calculations}, Springer, 2007.

\bibitem{ferguson2017bayeswham}
Ferguson, A.~L., {\em Journal of Computational Chemistry}, 2017, {\bf 38}(18),
  1583--1605.

\bibitem{kumar1992weighted}
Kumar, S.; Rosenberg, J.~M.; Bouzida, D.; Swendsen, R.~H. and Kollman, P.~A.,
  {\em Journal of Computational Chemistry}, 1992, {\bf 13}(8), 1011--1021.

\bibitem{bartels2000analyzing}
Bartels, C., {\em Chemical Physics Letters}, 2000, {\bf 331}(5), 446--454.

\bibitem{ferrenberg1989optimized}
Ferrenberg, A.~M. and Swendsen, R.~H., {\em Physical Review Letters}, 1989,
  {\bf 63}(12), 1195.

\bibitem{bennett1976efficient}
Bennett, C.~H., {\em Journal of Computational Physics}, 1976, {\bf 22}(2),
  245--268.

\bibitem{bartels1997multidimensional}
Bartels, C. and Karplus, M., {\em Journal of Computational Chemistry}, 1997,
  {\bf 18}(12), 1450--1462.

\bibitem{habeck2012bayesian}
Habeck, M., {\em Physical Review Letters}, 2012, {\bf 109}(10), 100601.

\bibitem{zhu2012convergence}
Zhu, F. and Hummer, G., {\em Journal of Computational Chemistry}, 2012, {\bf
  33}(4), 453--465.

\bibitem{gallicchio2005temperature}
Gallicchio, E.; Andrec, M.; Felts, A.~K. and Levy, R.~M., {\em The Journal of
  Physical Chemistry B}, 2005, {\bf 109}(14), 6722--6731.

\bibitem{hartmann2011two}
Hartmann, C.; Latorre, J.~C. and Ciccotti, G., {\em The European Physical
  Journal-Special Topics}, 2011, {\bf 200}(1), 73--89.

\bibitem{neidigh2002designing}
Neidigh, J.~W.; Fesinmeyer, R.~M. and Andersen, N.~H., {\em Nature Structural
  \& Molecular Biology}, 2002, {\bf 9}(6), 425.

\bibitem{sultan2017transfer}
Sultan, M.~M. and Pande, V.~S., {\em The Journal of Physical Chemistry B}, in
  press, 2017, page doi: 10.1021/acs.jpcb.7b06896.

\bibitem{edelsbrunner1983shape}
Edelsbrunner, H.; Kirkpatrick, D. and Seidel, R., {\em IEEE Transactions on
  Information Theory}, 1983, {\bf 29}(4), 551--559.

\bibitem{edelsbrunner1994three}
Edelsbrunner, H. and M{\"u}cke, E.~P., {\em ACM Transactions on Graphics},
  1994, {\bf 13}(1), 43--72.

\bibitem{edelsbrunner2003surface}
Edelsbrunner, H., {\em Algorithms and Combinatorics}, 2003, {\bf 25}, 379--404.

\bibitem{humphrey1996vmd}
Humphrey, W.; Dalke, A. and Schulten, K., {\em Journal of Molecular Graphics},
  1996, {\bf 14}(1), 33--38.

\bibitem{wang2004development}
Wang, J.; Wolf, R.~M.; Caldwell, J.~W.; Kollman, P.~A. and Case, D.~A., {\em
  Journal of Computational Chemistry}, 2004, {\bf 25}(9), 1157--1174.

\bibitem{schlick2010molecular}
Schlick, T., {\em Molecular modeling and simulation: an interdisciplinary
  guide: an interdisciplinary guide}, Vol. ~21, Springer Science \& Business
  Media, 2010.

\bibitem{hess1997lincs}
Hess, B.; Bekker, H.; Berendsen, H.~J. and Fraaije, J.~G., {\em Journal of
  Computational Chemistry}, 1997, {\bf 18}(12), 1463--1472.

\bibitem{allenoxford}
Allen, M. and Tildesley, D., {\em Computer Simulation of Liquids}, Oxford
  University Press, New York, 1987.

\bibitem{horn2004development}
Horn, H.~W.; Swope, W.~C.; Pitera, J.~W.; Madura, J.~D.; Dick, T.~J.; Hura,
  G.~L. and Head-Gordon, T., {\em The Journal of Chemical Physics}, 2004, {\bf
  120}(20), 9665--9678.

\bibitem{berman2002protein}
Berman, H.~M.; Battistuz, T.; Bhat, T.~N.; Bluhm, W.~F.; Bourne, P.~E.;
  Burkhardt, K.; Feng, Z.; Gilliland, G.~L.; Iype, L.; Jain, S.; Fagan, P.;
  Marvin, J.; Padilla, D.; Ravichandran, V.; Schneider, B.; Thanki, N.;
  Weissig, H.; Westbrook, J.~D. and Zardecki, C., {\em Acta Crystallographica
  Section D: Biological Crystallography}, 2002, {\bf 58}(6), 899--907.

\bibitem{andersen1980molecular}
Andersen, H.~C., {\em The Journal of Chemical Physics}, 1980, {\bf 72}(4),
  2384--2393.

\bibitem{chow1995isothermal}
Chow, K.-H. and Ferguson, D.~M., {\em Computer Physics Communications}, 1995,
  {\bf 91}(1-3), 283--289.

\bibitem{aaqvist2004molecular}
{\AA}qvist, J.; Wennerstr{\"o}m, P.; Nervall, M.; Bjelic, S. and Brandsdal,
  B.~O., {\em Chemical Physics Letters}, 2004, {\bf 384}(4), 288--294.

\bibitem{hockney1988computer}
Hockney, R.~W. and Eastwood, J.~W., {\em Computer Simulation Using Particles},
  CRC Press, 1988.

\bibitem{darden1993particle}
Darden, T.; York, D. and Pedersen, L., {\em The Journal of Chemical Physics},
  1993, {\bf 98}(12), 10089--10092.

\bibitem{2016arXiv160502688full}
Al-Rfou, R.; Alain, G.; Almahairi, A.; Angermueller, C.; Bahdanau, D.; Ballas,
  N.; Bastien, F.; Bayer, J.; Belikov, A.; Belopolsky, A.; Bengio, Y.;
  Bergeron, A.; Bergstra, J.; Bisson, V.; {Bleecher Snyder}, J.; Bouchard, N.;
  Boulanger-Lewandowski, N.; Bouthillier, X.; de~Br\'ebisson, A.; Breuleux, O.;
  Carrier, P.-L.; Cho, K.; Chorowski, J.; Christiano, P.; Cooijmans, T.;
  C\^ot\'e, M.-A.; C\^ot\'e, M.; Courville, A.; Dauphin, Y.~N.; Delalleau, O.;
  Demouth, J.; Desjardins, G.; Dieleman, S.; Dinh, L.; Ducoffe, M.; Dumoulin,
  V.; {Ebrahimi Kahou}, S.; Erhan, D.; Fan, Z.; Firat, O.; Germain, M.; Glorot,
  X.; Goodfellow, I.; Graham, M.; Gulcehre, C.; Hamel, P.; Harlouchet, I.;
  Heng, J.-P.; Hidasi, B.; Honari, S.; Jain, A.; Jean, S.; Jia, K.; Korobov,
  M.; Kulkarni, V.; Lamb, A.; Lamblin, P.; Larsen, E.; Laurent, C.; Lee, S.;
  Lefrancois, S.; Lemieux, S.; L\'eonard, N.; Lin, Z.; Livezey, J.~A.; Lorenz,
  C.; Lowin, J.; Ma, Q.; Manzagol, P.-A.; Mastropietro, O.; McGibbon, R.~T.;
  Memisevic, R.; van Merri\"enboer, B.; Michalski, V.; Mirza, M.; Orlandi, A.;
  Pal, C.; Pascanu, R.; Pezeshki, M.; Raffel, C.; Renshaw, D.; Rocklin, M.;
  Romero, A.; Roth, M.; Sadowski, P.; Salvatier, J.; Savard, F.; Schl\"uter,
  J.; Schulman, J.; Schwartz, G.; Serban, I.~V.; Serdyuk, D.; Shabanian, S.;
  Simon, E.; Spieckermann, S.; Subramanyam, S.~R.; Sygnowski, J.; Tanguay, J.;
  van Tulder, G.; Turian, J.; Urban, S.; Vincent, P.; Visin, F.; de~Vries, H.;
  Warde-Farley, D.; Webb, D.~J.; Willson, M.; Xu, K.; Xue, L.; Yao, L.; Zhang,
  S. and Zhang, Y., {\em arXiv e-prints}, 2016, {\bf
  http://arxiv.org/abs/1605.02688}.

\bibitem{hashemian2015topological}
Hashemian, B. and Arroyo, M., {\em The Journal of Chemical Physics}, 2015, {\bf
  142}(4), 044102.

\bibitem{chodera2007use}
Chodera, J.~D.; Swope, W.~C.; Pitera, J.~W.; Seok, C. and Dill, K.~A., {\em
  Journal of Chemical Theory and Computation}, 2007, {\bf 3}(1), 26--41.

\bibitem{hummer2003coarse}
Hummer, G. and Kevrekidis, I.~G., {\em The Journal of Chemical Physics}, 2003,
  {\bf 118}(23), 10762--10773.

\bibitem{chodera2006long}
Chodera, J.~D.; Swope, W.~C.; Pitera, J.~W. and Dill, K.~A., {\em Multiscale
  Modeling \& Simulation}, 2006, {\bf 5}(4), 1214--1226.

\bibitem{chodera2007automatic}
Chodera, J.~D.; Singhal, N.; Pande, V.~S.; Dill, K.~A. and Swope, W.~C., {\em
  The Journal of Chemical Physics}, 2007, {\bf 126}(15), 04B616.

\bibitem{bolhuis2000reaction}
Bolhuis, P.~G.; Dellago, C. and Chandler, D., {\em Proceedings of the National
  Academy of Sciences}, 2000, {\bf 97}(11), 5877--5882.

\bibitem{jakli2012variation}
J{\'a}kli, I.; Jensen, S.~J.~K.; Csizmadia, I.~G. and Perczel, A., {\em
  Chemical Physics Letters}, 2012, {\bf 547}, 82--88.

\bibitem{patrangenaru2015nonparametric}
Patrangenaru, V. and Ellingson, L., {\em Nonparametric Statistics on Manifolds
  and Their Applications to Object Data Analysis}, CRC Press, 2015.

\bibitem{whitney1936differentiable}
Whitney, H., {\em Annals of Mathematics}, 1936, pages 645--680.

\bibitem{broomhead1986extracting}
Broomhead, D.~S. and King, G.~P., {\em Physica D: Nonlinear Phenomena}, 1986,
  {\bf 20}(2-3), 217--236.

\bibitem{kim2015systematic}
Kim, S.~B.; Dsilva, C.~J.; Kevrekidis, I.~G. and Debenedetti, P.~G., {\em The
  Journal of Chemical Physics}, 2015, {\bf 142}(8), 085101.

\bibitem{qiu2002smaller}
Qiu, L.; Pabit, S.~A.; Roitberg, A.~E. and Hagen, S.~J., {\em Journal of the
  American Chemical Society}, 2002, {\bf 124}(44), 12952--12953.

\bibitem{seshasayee2005high}
Seshasayee, A.~S.~N., {\em Theoretical Biology and Medical Modelling}, 2005,
  {\bf 2}(1), 7.

\bibitem{heyda2011urea}
Heyda, J.; Kozisek, M.; Bedn{\'a}rova, L.; Thompson, G.; Konvalinka, J.;
  Vondrasek, J. and Jungwirth, P., {\em The Journal of Physical Chemistry B},
  2011, {\bf 115}(28), 8910--8924.

\bibitem{juraszek2006sampling}
Juraszek, J. and Bolhuis, P., {\em Proceedings of the National Academy of
  Sciences}, 2006, {\bf 103}(43), 15859--15864.

\bibitem{kim2016computational}
Kim, S.~B.; Gupta, D.~R. and Debenedetti, P.~G., {\em Scientific Reports},
  2016, {\bf 6}, 25612.

\bibitem{hatch2014computational}
Hatch, H.~W.; Stillinger, F.~H. and Debenedetti, P.~G., {\em The Journal of
  Physical Chemistry B}, 2014, {\bf 118}(28), 7761--7769.

\bibitem{kannan2014role}
Kannan, S. and Zacharias, M., {\em PloS ONE}, 2014, {\bf 9}(2), e88383.

\bibitem{schug2005energy}
Schug, A.; Wenzel, W. and Hansmann, U.~H., {\em The Journal of Chemical
  Physics}, 2005, {\bf 122}(19), 194711.

\bibitem{snow2002trp}
Snow, C.~D.; Zagrovic, B. and Pande, V.~S., {\em Journal of the American
  Chemical Society}, 2002, {\bf 124}(49), 14548--14549.

\bibitem{zhou2003trp}
Zhou, R., {\em Proceedings of the National Academy of Sciences}, 2003, {\bf
  100}(23), 13280--13285.

\bibitem{zhou2003free}
Zhou, R., {\em Proteins: Structure, Function, and Bioinformatics}, 2003, {\bf
  53}(2), 148--161.

\bibitem{deng2013kinetics}
Deng, N.-j.; Dai, W. and Levy, R.~M., {\em The Journal of Physical Chemistry
  B}, 2013, {\bf 117}(42), 12787--12799.

\bibitem{patriksson2007direct}
Patriksson, A.; Adams, C.~M.; Kjeldsen, F.; Zubarev, R.~A. and van~der Spoel,
  D., {\em The Journal of Physical Chemistry B}, 2007, {\bf 111}(46),
  13147--13150.

\bibitem{pitera2003understanding}
Pitera, J.~W. and Swope, W., {\em Proceedings of the National Academy of
  Sciences}, 2003, {\bf 100}(13), 7587--7592.

\bibitem{meuzelaar2013folding}
Meuzelaar, H.; Marino, K.~A.; Huerta-Viga, A.; Panman, M.~R.; Smeenk, L.~E.;
  Kettelarij, A.~J.; van Maarseveen, J.~H.; Timmerman, P.; Bolhuis, P.~G. and
  Woutersen, S., {\em The Journal of Physical Chemistry B}, 2013, {\bf
  117}(39), 11490--11501.

\bibitem{barua2008trp}
Barua, B.; Lin, J.~C.; Williams, V.~D.; Kummler, P.; Neidigh, J.~W. and
  Andersen, N.~H., {\em Protein Engineering, Design \& Selection}, 2008, {\bf
  21}(3), 171--185.

\bibitem{iavarone2005conformational}
Iavarone, A.~T. and Parks, J.~H., {\em Journal of the American Chemical
  Society}, 2005, {\bf 127}(24), 8606--8607.

\bibitem{rovo2013structural}
Rov{\'o}, P.; Str{\'a}ner, P.; L{\'a}ng, A.; Bartha, I.; Husz{\'a}r, K.;
  Nyitray, L. and Perczel, A., {\em Chemistry-A European Journal}, 2013, {\bf
  19}(8), 2628--2640.

\bibitem{adams2006probing}
Adams, C.~M.; Kjeldsen, F.; Patriksson, A.; van Der~Spoel, D.; Gr{\"a}slund,
  A.; Papadopoulos, E. and Zubarev, R.~A., {\em International Journal of Mass
  Spectrometry}, 2006, {\bf 253}(3), 263--273.

\bibitem{bolhuis2009two}
Bolhuis, P.~G., {\em Frontiers in Bioscience}, 2009, {\bf 14}, 2801--2828.

\bibitem{kannan2009folding}
Kannan, S. and Zacharias, M., {\em Proteins: Structure, Function, and
  Bioinformatics}, 2009, {\bf 76}(2), 448--460.

\bibitem{shao2012enhanced}
Shao, Q.; Shi, J. and Zhu, W., {\em The Journal of Chemical Physics}, 2012,
  {\bf 137}(12), 125103.

\bibitem{paschek2008computing}
Paschek, D.; Hempel, S. and Garc{\'\i}a, A.~E., {\em Proceedings of the
  National Academy of Sciences}, 2008, {\bf 105}(46), 17754--17759.

\bibitem{kaminski2001evaluation}
Kaminski, G.~A.; Friesner, R.~A.; Tirado-Rives, J. and Jorgensen, W.~L., {\em
  The Journal of Physical Chemistry B}, 2001, {\bf 105}(28), 6474--6487.

\bibitem{lindahl2001gromacs}
Lindahl, E.; Hess, B. and Van Der~Spoel, D., {\em Journal of Molecular
  Modeling}, 2001, {\bf 7}(8), 306--317.

\bibitem{ahmed2005uv}
Ahmed, Z.; Beta, I.~A.; Mikhonin, A.~V. and Asher, S.~A., {\em Journal of the
  American Chemical Society}, 2005, {\bf 127}(31), 10943--10950.

\bibitem{best2008current}
Best, R.~B.; Buchete, N.-V. and Hummer, G., {\em Biophysical Journal}, 2008,
  {\bf 95}(1), L07--L09.

\bibitem{best2010protein}
Best, R.~B. and Mittal, J., {\em The Journal of Physical Chemistry B}, 2010,
  {\bf 114}(46), 14916--14923.

\bibitem{abascal2005general}
Abascal, J.~L. and Vega, C., {\em The Journal of Chemical Physics}, 2005, {\bf
  123}(23), 234505.

\bibitem{cornell1995second}
Cornell, W.~D.; Cieplak, P.; Bayly, C.~I.; Gould, I.~R.; Merz, K.~M.; Ferguson,
  D.~M.; Spellmeyer, D.~C.; Fox, T.; Caldwell, J.~W. and Kollman, P.~A., {\em
  Journal of the American Chemical Society}, 1995, {\bf 117}(19), 5179--5197.

\bibitem{jorgensen1983comparison}
Jorgensen, W.~L.; Chandrasekhar, J.; Madura, J.~D.; Impey, R.~W. and Klein,
  M.~L., {\em The Journal of Chemical Physics}, 1983, {\bf 79}(2), 926--935.

\bibitem{lindorff2011fast}
Lindorff-Larsen, K.; Piana, S.; Dror, R.~O. and Shaw, D.~E., {\em Science},
  2011, {\bf 334}(6055), 517--520.

\bibitem{mackerell1998all}
MacKerell, A.~D.; Bashford, D.; Bellott, M.; Dunbrack, R.~L.; Evanseck, J.~D.;
  Field, M.~J.; Fischer, S.; Gao, J.; Guo, H.; Ha, S.; Joseph-McCarthy, D.;
  Kuchnir, L.; Kuczera, K.; Lau, F.~T.~K.; Mattos, C.; Michnick, S.; Ngo, T.;
  Nguyen, D.~T.; Prodhom, B.; Reiher, W.~E.; Roux, B.; Schlenkrich, M.; Smith,
  J.~C.; Stote, R.; Straub, J.; Watanabe, M.; Wi{\'o}rkiewicz-Kuczera, J.; Yin,
  D. and Karplus, M., {\em The Journal of Physical Chemistry B}, 1998, {\bf
  102}(18), 3586--3616.

\bibitem{mackerell2004extending}
MacKerell, A.~D.; Feig, M. and Brooks, C.~L., {\em Journal of Computational
  Chemistry}, 2004, {\bf 25}(11), 1400--1415.

\bibitem{kmiecik2016coarse}
Kmiecik, S.; Gront, D.; Kolinski, M.; Wieteska, L.; Dawid, A.~E. and Kolinski,
  A., {\em Chemical Reviews}, 2016, {\bf 116}(14), 7898--7936.

\end{thebibliography}


\clearpage










\end{document}